\documentclass[pre,aps,twocolumn,superscriptaddress,longbibliography]{revtex4-2}
\usepackage{amsmath}
\usepackage{amssymb} 
\usepackage{bm}
\usepackage{color}
\usepackage{commath}
\usepackage{dcolumn}
\usepackage{graphicx}
\usepackage{hyperref} 
\usepackage{multirow}
\usepackage{subfigure}

\begin{document}

\title{Planted vertex cover problem on regular random graphs and nonmonotonic temperature-dependence in the supercooled region}

\author{Xin-Yi Fan}

\affiliation{
  Institute of Theoretical Physics, Chinese Academy of Sciences, Beijing 100190, China}

\affiliation{
  School of Physical Sciences, University of Chinese Academy of Sciences, Beijing 100049, China}

\author{Hai-Jun Zhou}
\email{zhouhj@itp.ac.cn}

\affiliation{
  Institute of Theoretical Physics, Chinese Academy of Sciences, Beijing 100190, China}

\affiliation{
  School of Physical Sciences, University of Chinese Academy of Sciences, Beijing 100049, China}

\affiliation{
  MinJiang Collaborative Center for Theoretical Physics, MinJiang University, Fuzhou 350108, China
}

\date{\today}

\begin{abstract}
  We introduce a planted vertex cover problem on regular random graphs and study it by the cavity method of statistical mechanics. Different from conventional Ising models, the equilibrium ferromagnetic phase transition of this binary-spin two-body interaction system is discontinuous, as the paramagnetic phase is separated from the ferromagnetic phase by an extensive free energy barrier. The free energy landscape can be distinguished into three different types depending on the two degree parameters of the planted graph. The critical inverse temperatures at which the paramagnetic phase becomes locally unstable towards the ferromagnetic phase ($\beta_{\textrm{pf}}$) and towards spin glass phases ($\beta_{\textrm{pg}}$) satisfy $\beta_{\textrm{pf}} > \beta_{\textrm{pg}}$, $\beta_{\textrm{pf}} < \beta_{\textrm{pg}}$ and $\beta_{\textrm{pf}} = \beta_{\textrm{pg}}$, respectively, in these three landscapes. A locally stable anti-ferromagnetic phase emerges in the free energy landscape if $\beta_{\textrm{pf}} < \beta_{\textrm{pg}}$. When exploring the free energy landscape by stochastic local search dynamics, we find that in agreement with our theoretical prediction, the first-passage time from the paramagnetic phase to the ferromagnetic phase is nonmonotonic with the inverse temperature. The potential relevance of the planted vertex cover model to supercooled glass-forming liquids is briefly discussed.
\end{abstract}

\maketitle

\section{Introduction}

Planted sparse random-graph optimization problems at the interface between statistical physics and statistical inference have rich structures in their free energy landscapes~\cite{Zdeborova-Krzakala-2016,Mezard-Montanari-2009}. Some widely studied examples include low-density parity-check codes~\cite{Gallager-1962,Kabashima-Saad-2004}, planted $p$-spin models and the Sourlass code~\cite{Sourlas-1989,Franz-etal-2001,Huang-Zhou-2009}, planted $K$-satisfiability~\cite{Barthel-etal-2002,Li-Ma-Zhou-2009,Krzakala-Mezard-Zdeborova-2014}, planted coloring and the Potts model~\cite{Krzakala-Zdeborova-2009,Zhou-2019,Angelini-RicciTersenghi-2022}. These planted problems are dictated by many-body interactions and/or are composed of vertices with multiple discrete states, which often lead to an equilibrium discontinuous phase transition between the paramagnetic phase of disordered configurations and the phase of the ordered configurations informative of the planted ground state. Disordered spin glass phases are also present in these planted graphical problems at high enough inverse temperatures and they further increase the computational difficulty of retrieving the planted ground state. 

In the present work, we introduce a planted version of the vertex cover problem and study it by the cavity method of statistical physics~\cite{Zhou-2003a,Weigt-Zhou-2006,Zhao-Zhou-2014,Zhou-2015}. The vertex cover problem is a fundamental combinatorial optimization problem involving only two-body interactions and only binary vertex states. Each vertex is either being occupied or being empty and each edge prohibits the two incident vertices from being simultaneously empty~\cite{Weigt-Hartmann-2000,Hartmann-Weigt-2003}. Unlike the continuous ferromagnetic phase transition in conventional Ising models, we find that the equilibrium ferromagnetic phase transition of the planted vertex cover system is discontinuous in nature. This result demonstrates that many-body interactions and multiple discrete states both are \emph{not} necessary conditions for discontinuous ferromagnetic phase transitions in a planted optimization problem. In addition, we find that the free energy landscape of the planted vertex cover problem has other novel features. One dynamic consequence of the free energy landscape structure is the nonmonotonic behavior of the mean first-passage time as a function of inverse temperature $\beta$.

\begin{figure}[b]
  \centering
  \subfigure[]{
    \label{fig:PRR}
    \includegraphics[width=0.465\linewidth]{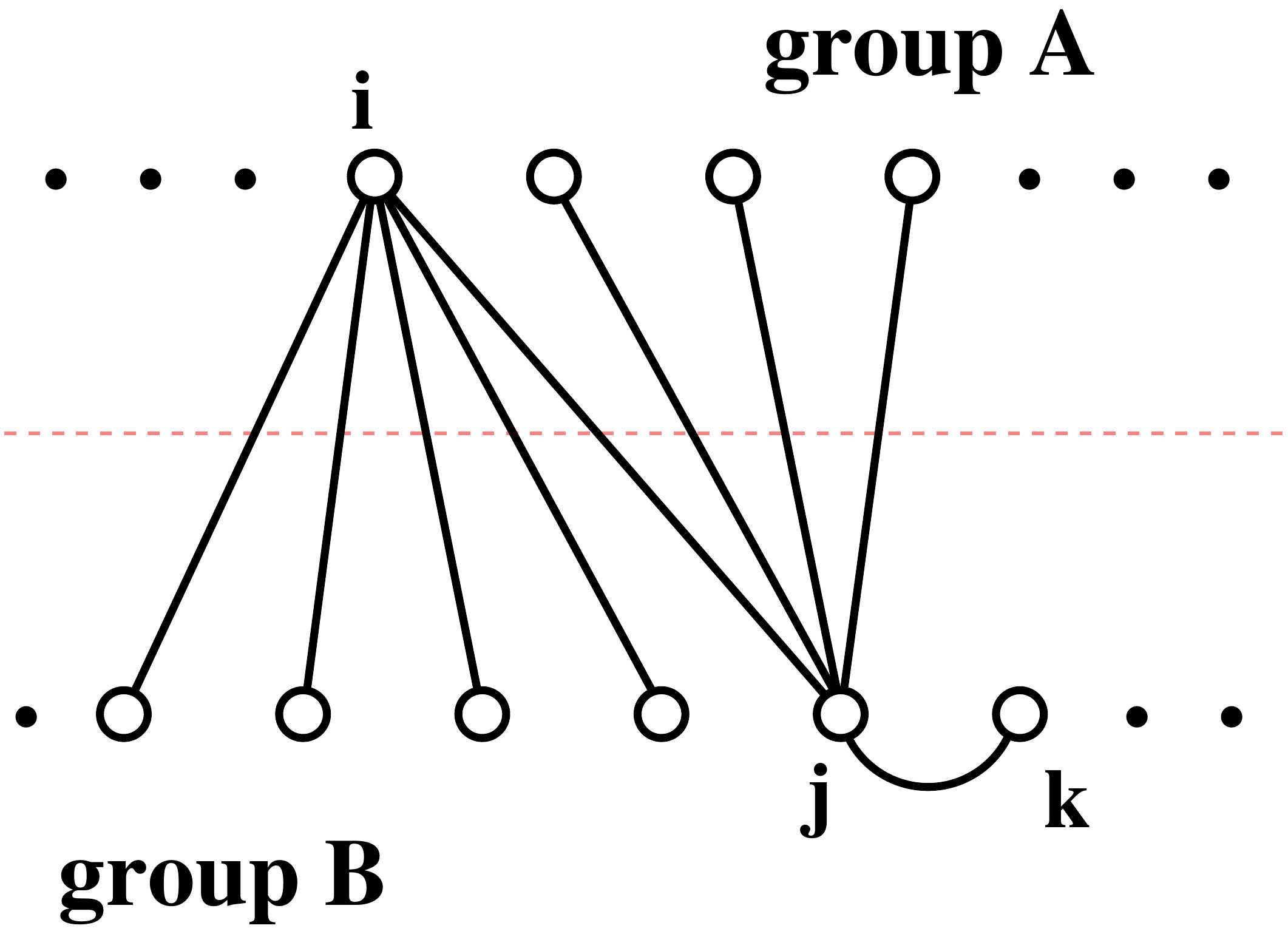}
  }
  \subfigure[]{
    \label{fig:LSt}
    \includegraphics[width=0.465\linewidth]{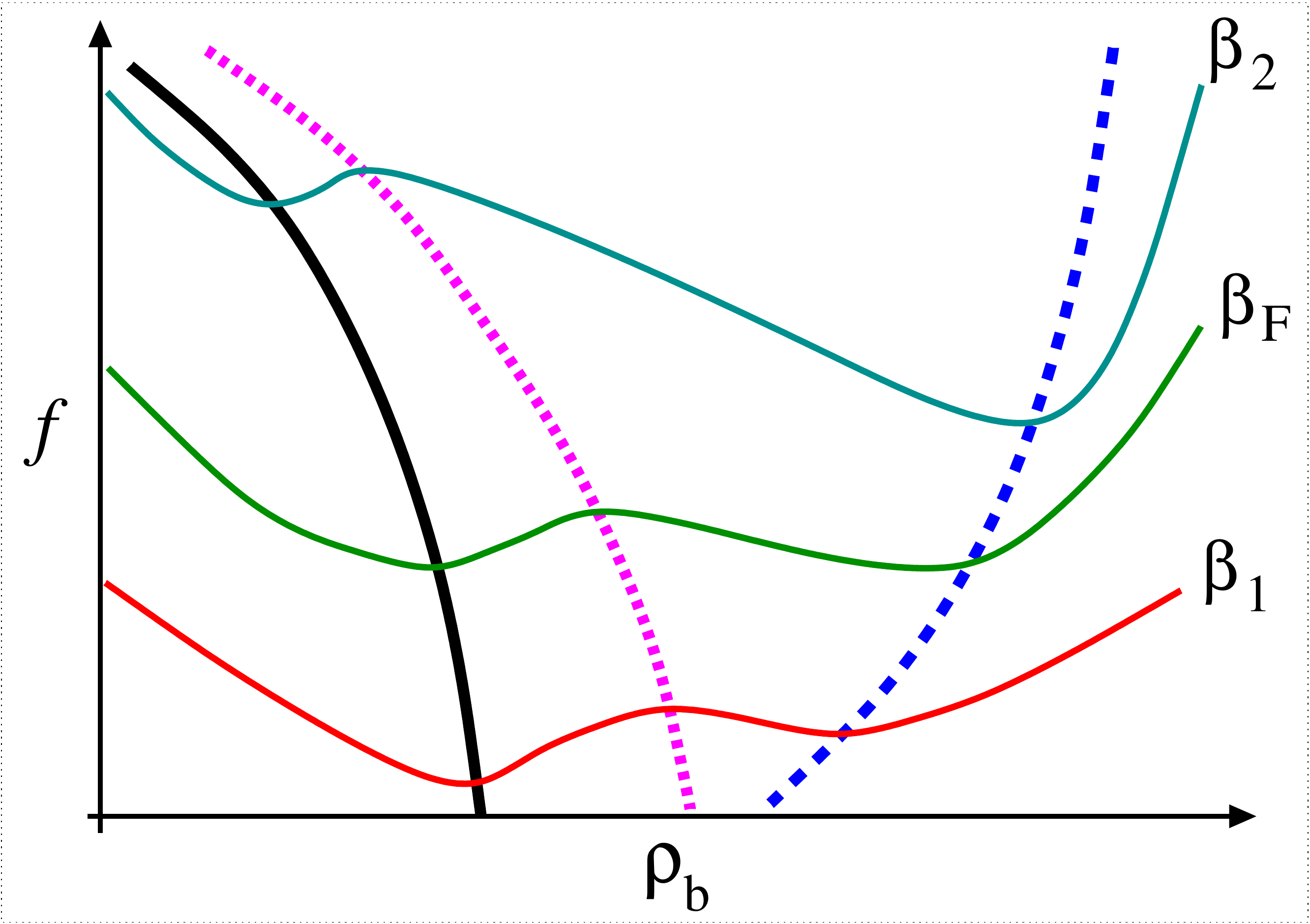}
  }
  
  \subfigure[]{
    \label{fig:LUns}
    \includegraphics[width=0.465\linewidth]{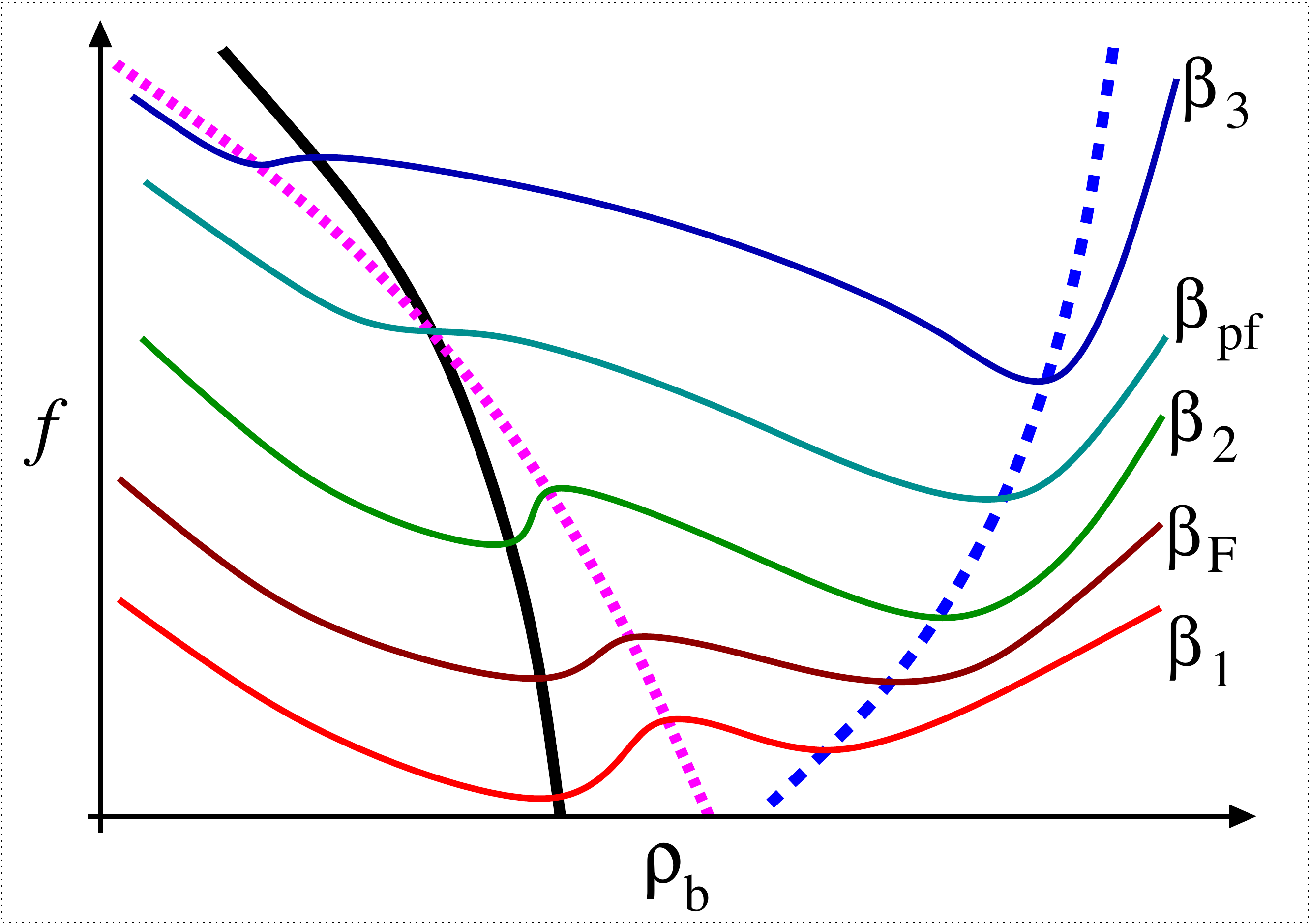}
  }
  \subfigure[]{
    \label{fig:LMr}
    \includegraphics[width=0.465\linewidth]{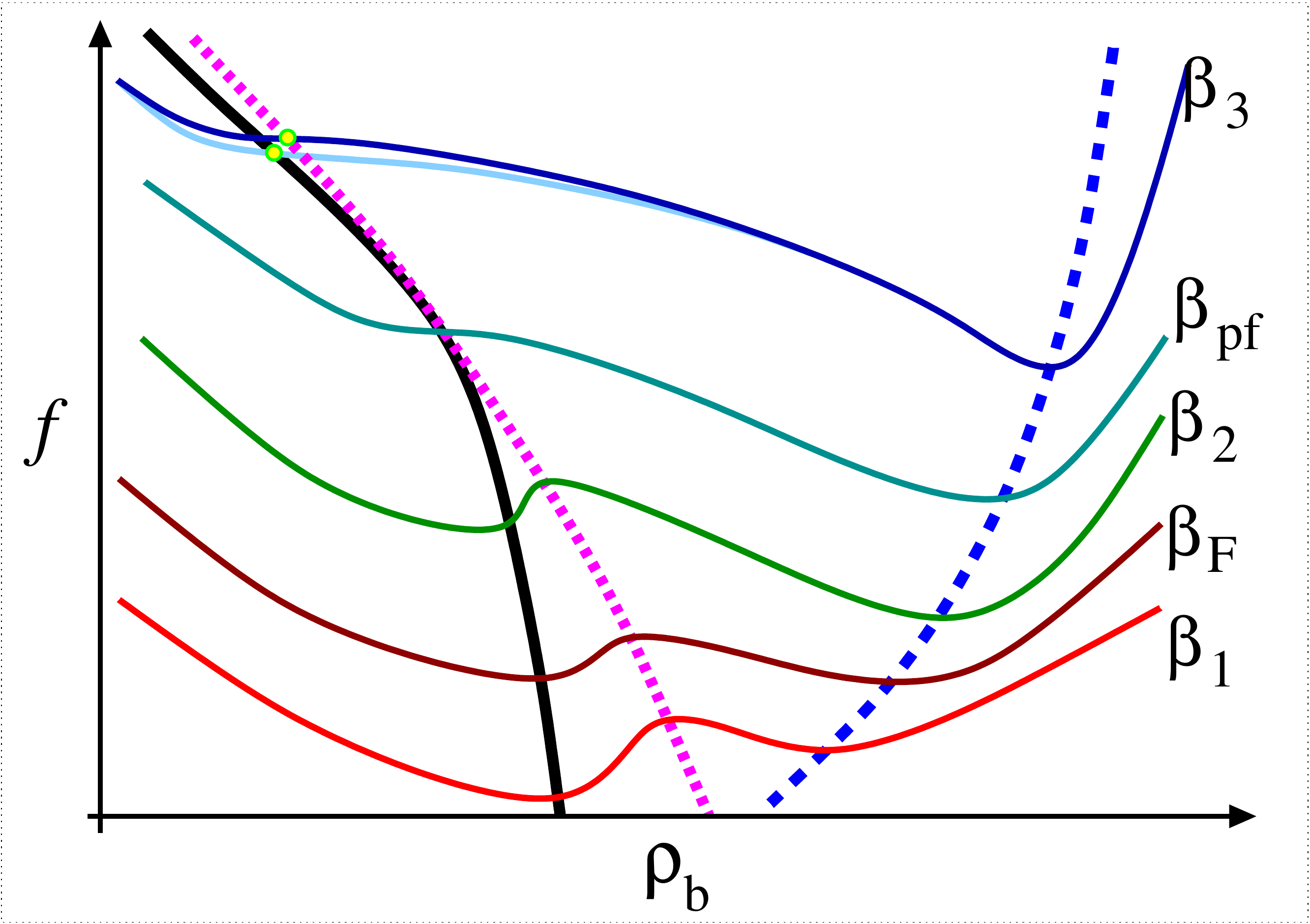}
  }
  \caption{
    \label{fig:landscape}
    Free energy landscape of the planted vertex cover problem. (a) Regular random graph containing two groups of vertices and free of edges within group $A$. (b)-(d) Schematic plots of free energy density $f$ versus the fraction $\rho_{\textrm{b}}$ of occupied vertices in group $B$ at different fixed inverse temperatures ($\beta_1 < \beta_F < \beta_2 < \beta_{\textrm{pf}} < \beta_3$). The three phases DS, MP and CP are respectively indicated by the solid, dotted, and dashed thick lines. The equilibrium discontinuous phase transition occurs at $\beta_{F}$. (b) DS phase is locally stable. (c) DS phase is locally unstable when inverse temperature exceeds $\beta_{\textrm{pf}}$. (d) DS and MP phases touch at $\beta_{\textrm{pf}}$, and at $\beta_3 > \beta_{\textrm{pf}}$ both the DS and MP phases (indicated by small circles) are saddle points and they can evolve to the CP phase without passing through each other.
  }
  \end{figure}

The vertices in the underlying random graph of our planted problem form two different groups, $A$ and $B$, such that there is no edge between any two vertices of group $A$ (Fig.~\ref{fig:PRR}). The graph is regular in the sense that all the vertices have the same number of nearest neighboring vertices. The mean field belief-propagation (BP) equation of such a planted random graph ensemble has three different fixed points, which correspond to the paramagnetic disordered symmetric (DS) phase, the ferromagnetic canonically polarized (CP) phase, and another intermediate microcanonically polarized (MP) phase, respectively. (In this work we follow the notation of Refs.~\cite{Zhou-2019,Zhou-Liao-2021} and refer to the three fixed points as DS, CP, and MP.) By computing the free energy densities and analyzing the local stability properties of these fixed points at different values of the inverse temperature $\beta$, we infer that the free energy landscape can be classified into three different types. We offer concrete examples of random graph ensembles to help appreciating these distinct landscapes. 

Figure~\ref{fig:LSt} is a schematic plot of one common type of free energy landscape with two valleys separated by a barrier. One valley is the paramagnetic DS phase. The statistical properties of the vertices of groups $A$ and $B$ can not be distinguished in this phase; for example, the marginal probabilities $\rho_{\textrm{a}}$ and $\rho_{\textrm{b}}$ of being occupied are the same for vertices in groups $A$ and $B$, so the magnetization order parameter $m$ ($\equiv \rho_{\textrm{b}}-\rho_{\textrm{a}}$) is zero. The other valley is the ferromagnetic CP phase, which are formed by microscopic configurations that are very similar to the planted ground state, with $\rho_{\textrm{b}}$ much higher than $\rho_{\textrm{a}}$ and hence large magnetization $m$. The MP fixed point of the BP equation describes the free energy barrier; the microscopic configurations of this unstable phase are partially polarized ($\rho_{\textrm{b}} > \rho_{\textrm{a}}$) and hence have intermediate positive magnetization $m$. The line of MP fixed points in Fig.~\ref{fig:LSt} is the watershed separating the DS and the CP free energy valleys.

For this type of free energy landscape, an unexpected prediction of the mean field theory is that there exists an optimal inverse temperature value $\beta_{opt}$ at which the escaping rate from the paramagnetic DS phase to the ferromagnetic CP phase achieves the maximum value (Fig.~\ref{fig:K10v5v5diff}). This nonmonotonic relaxation property is absent in the planted $p$-spin interaction model~\cite{Zhou-Liao-2021,Bellitti-etal-2021}, even though it has a qualitatively similar free energy landscape as Fig.~\ref{fig:LSt}.

Figure~\ref{fig:LUns} is a schematic plot of another common type of free energy landscape. It is distinguished from Fig.~\ref{fig:LSt}) in that the lines of DS and MP fixed points cross with each other at an inverse temperature value $\beta_{\textrm{pf}}$, which also marks the critical point beyond which the paramagnetic DS phase becomes locally unstable. The free energy density of the MP phase becomes lower than that of the DS phase at $\beta > \beta_{\textrm{pf}}$ and its magnetization changes from being positive to being negative, indicating the emergence of a stable anti-ferromagnetic phase at $\beta > \beta_{\textrm{pf}}$. This anti-ferromagnetic MP free energy valley is separated from the ferromagnetic CP valley by the paramagnetic DS phase which is now a free energy barrier.

Besides these two common scenarios, the free energy landscape may also take the peculiar form of Fig.~\ref{fig:LMr} for some special planted regular random graph ensembles. One major distinctive feature of Fig.~\ref{fig:LMr} is that the lines of MP and DS fixed points do not cross with each other but touch at the critical inverse temperature $\beta_{\textrm{pf}}$. The paramagnetic DS phase has lower free energy density than the MP phase and therefore it is locally stable with respect to the MP phase at all inverse temperatures, except for the single point $\beta_{\textrm{pf}}$ at which the DS and MP phases merge together. On the other hand, we find that at $\beta > \beta_{\textrm{pf}}$ the paramagnetic DS phase is no longer locally stable with respect to the ferromagnetic CP phase. Both the DS and the MP fixed points are saddle points of the free energy landscape at $\beta > \beta_{\textrm{pf}}$ and only the ferromagnetic CP phase is stable. Furthermore, at the critical point $\beta_{\textrm{pf}}$ the paramagnetic DS phase also becomes locally unstable towards disordered spin glass phases. In other words, $\beta_{\textrm{pf}} = \beta_{\textrm{pg}}$ where $\beta_{\textrm{pg}}$ is the critical inverse temperature at which the paramagnetic DS phase becomes locally unstable towards disordered spin glass states.

We perform Markov-Chain Monte Carlo simulations with local updating rules obeying the detailed balance condition. Our numerical results obtained on finite systems demonstrate that, when the inverse temperature $\beta$ is located slightly below the theoretically predicted value $\beta_{opt}$ (systems of type Fig.~\ref{fig:LSt}) or $\beta_{\textrm{pf}}$ (systems of type \ref{fig:LMr}),  it is most efficient for the stochastic local search dynamics to escaping from the paramagnetic DS phase to the ferromagnetic CP phase. 

This work help us gain new insights into the free energy landscape of planted random vertex cover problems. The work may also be relevant to other planted random optimization and inference problems. The planted regular random graph ensembles studied here may serve as two-body interaction benchmark problems for quantum computing algorithms~\cite{Bellitti-etal-2021,Zeng-etal-2024}. The vertex cover problem is also a special case of lattice glass models~\cite{Biroli-Mezard-2002,Rivoire-etal-2004}, which contain geometrical frustrations such that an empty vertex can have at most $z$ empty nearest neighbors ($z = 0$ for the vertex cover problem). Because there is a planted configuration, the planted vertex cover problem can be considered as a model of lattice glass with a crystalline ground state or as a graphical model of glass-forming liquids. The planted vertex cover problem can also be considered as a kinetically severely constrained spin system with a planted crystalline state~\cite{Zhou-2024}. The theoretical and simulation results presented in this work, including the nonmonotonic free energy and entropy barriers and nonmonotonic relaxation dynamics, may be indicative of some general nonequilibrium properties of supercooled glass-forming liquids~\cite{Schmelzer-etal-2016,Jeng-2006,Zhang-Hou-2022} and kinetically constrained spin models~\cite{Zhou-2024}.

We introduce the planted vertex cover problem in Sec.~\ref{sec:pvc} and outline the mean field theory in Sec.~\ref{sec:mct}. We then discuss the three types of free energy landscapes through conrete example systems in Secs.~\ref{sec:10_5_5}, \ref{sec:10_7_3} and \ref{sec:special}. The critical inverse temperatures $\beta_{\textrm{pf}}$ and $\beta_{\textrm{pg}}$ are analytically determined through local stability analysis in Sec.~\ref{sec:lsa}. A continuous-time BP evolution dynamics is studied in Sec.~\ref{sec:bpcted} to further check the local stability property of the free energy landscape. Finally we conclude this work in Sec.~\ref{sec:co} and mention some possible future extensions.

\section{Planted vertex cover problem}
\label{sec:pvc}

The vertex cover problem is defined on an undirected graph $G$ which contains $N$ vertices and $M$ edges. Each vertex $i$ has a binary occupation state, $c_i = 0$ (empty) or $c_i = 1$ (occupied). Each edge $(i, j)$ between two vertices $i$ and $j$ brings a hard constraint
\begin{equation}
  c_i + c_j \, \geq \, 1
  \label{eq240919a}
\end{equation}
to the states $c_i$ and $c_j$, which means that at least one incident vertex must be occupied. This hard constraint can also be expressed in the form of an equality as
$$
c_i + c_j - c_i \, c_j \, = \, 1 \; .
$$
If a microscopic occupation configuration of the whole system satisfies all the $M$ edge constraints, the set of all the occupied vertices,
$$
\Gamma\, \equiv \, \{i: \, \, c_i = 1\} \; ,
$$
is then said to be a vertex cover for the graph~\cite{Weigt-Hartmann-2000,Hartmann-Weigt-2003}. The complement of a vertex cover is referred to as an independent set, whose vertices are not adjacent to each other. The vertex cover (independent set) problem is a basic nondeterministic polynomial hard optimization problem in computer science with broad applications~\cite{Hartmann-Weigt-2003,Zhao-Zhou-2014,Benson-Kleinberg-2018,Zhao-etal-2020}. It can also be interpreted as a lattice glass model on a graph or as a special example of kinetically constrained spin systems~\cite{Biroli-Mezard-2002,Zhou-2024}. 

We consider a regular random graph $G$ which is formed by two groups of vertices (Fig.~\ref{fig:PRR}). There are $N_{\textrm{a}}$ vertices in group $A$ and $N_{\textrm{b}}$ ($=N-N_{\textrm{a}}$) vertices in group $B$. Each vertex of group $A$ is connected by $K$ undirected edges to $K$ randomly chosen vertices of group $B$. There is no edge between any two vertices of group $A$. Each vertex of group $B$ is also connected to $K$ other vertices, but only $K_{\textrm{ba}}$ of these nearest neighbors are belonging to group $A$ and the remaining $K_{\textrm{bb}}$ ($= K - K_{\textrm{ba}}$) vertices are members of group $B$.  The edge connection patterns of the graph are otherwise completely random. In the thermodynamic limit of $N\rightarrow \infty$, the statistical property of the graph is fully determined by the two degree parameters $K$ and $K_{\textrm{ba}}$. By counting the total number of inter-group edges, we get $N_{\textrm{a}} K = N_{\textrm{b}} K_{\textrm{ba}}$  which means
\begin{equation}
  \frac{N_{\textrm{a}}}{N} \, = \, \frac{K_{\textrm{ba}}}{K + K_{\textrm{ba}}} \; ,
  \quad \quad 
  \frac{N_{\textrm{b}}}{N} \, = \, \frac{K}{K + K_{\textrm{ba}}} \; .
\end{equation}

If the cardinality of a vertex cover achieves the minimum value among all the possible vertex covers of the graph, we say that it is a minimum vertex cover. Notice that the planted regular random graph has a planted vertex cover $\Gamma_0 = B$, and the vertex set $A$ is a planted independent set. The cardinality of $\Gamma_0$ is $N_{\textrm{b}}$ and the corresponding fraction of occupied vertices is $\rho_0 = K / (K + K_{\textrm{ba}})$. The relative cardinality (with respect to graph size $N$) of the minimum vertex cover solutions of the planted regular random graph must then be upper-bounded by $\rho_0$.

If the inter-group degree $K_{\textrm{ba}}$ to satisfy $2 \leq K_{\textrm{ba}} < K$, in the thermodynamic limit $N\rightarrow \infty$, we expect that the set $\Gamma_0$ will be the unique minimum vertex cover solution (ground state) with probability approaching unity. Although we have not yet verified this by rigorous mathematical analysis, the mean field theory of the next section predicts that $\rho_0$ is the ensemble-averaged minimum density of occupied vertices for this planted vertex cover problem. For our present work, whether the vertex set $B$ is really a minimum vertex cover solution is not essential. The key point is that there exists at least one vertex cover solution with relative cardinality at most $\rho_0$.

\section{Mean field cavity theory}
\label{sec:mct}

We define the energy of a microscopic occupation configuration $\vec{\bm{c}} = (c_1, c_2, \ldots, c_N)$ of the graph $G$ as
\begin{equation}
  E(\vec{\bm{c}}) \, = \, \sum\limits_{i=1}^{N} c_i \; ,
\end{equation}
which counts the total number of occupied vertices. We introduce an inverse temperature $\beta$ as the conjugate of $E(\vec{\bm{c}})$ and write down the partition function $Z(\beta)$ as
\begin{equation}
  Z(\beta) \, = \, \sum\limits_{\vec{\bm{c}}} \prod\limits_{i=1}^N e^{- \beta c_i}
  \prod\limits_{(j, k) \in G} \bigl( c_j + c_k - c_j \, c_k \bigr ) \; .
  \label{eq:Z}
\end{equation}  

It may be more suitable to call the control parameter $\beta$ the chemical potential since the conjugate quantity is the number of occupied vertices. Here we call $\beta$ the inverse temperature just to follow the convention of spin glass literature.

If configuration $\vec{\bm{c}}$ is a valid vertex cover, it contributes a term $e^{- \beta E(\vec{\bm{c}})}$ to the partition function, but if $\vec{\bm{c}}$ violates one or more of the edge constraints its contribution to $Z(\beta)$ vanishes. Note that as the inverse temperature $\beta$ increases the contributions of the lower-energy and minimum vertex cover solutions become more and more significant. At the limit of $\beta\rightarrow \infty$ only the minimum vertex cover solutions are relevant for the partition function.

\subsection{Belief-propagation equation}

The vertex cover partition function (\ref{eq:Z}) has been well studied by the mean field cavity method~\cite{Zhao-Zhou-2014}. Here we briefly review this theory at the replica-symmetric level. A key observation is that short loops are very rare in random graphs, and for such graphs when the inverse temperature $\beta$ is not too large the states of the nearest neighbors of a vertex $i$ will become independent if we delete vertex $i$ from the graph and wait until the remaining system recovers equilibrium~\cite{Mezard-Montanari-2009,Zhou-2015}. Let us denote by $\partial i \equiv \{ j : (i, j) \in G\}$ the vertex set containing all the nearest neighbors of $i$. For a vertex $j$ in this neighbor set $\partial i$, denote by $q_{j\rightarrow i}$ its equilibrium probability of being occupied ($c_j = 1$) in the absence of vertex $i$. Using the conditional independence assumption (known in the literature as the Bethe-Peierls approximation) we can write down the following self-consistent belief-propagation (BP) equation
\begin{equation}
  \label{eq:BPoriginal}
  q_{j \rightarrow i} \, = \,  \frac{ e^{-\beta} }{ e^{-\beta} +
    \prod\limits_{k\in \partial j \backslash i} q_{k \rightarrow j}} \; ,
\end{equation}
where $\partial j \backslash i$ contains all the other nearest neighbors of vertex $j$ except vertex $i$. The whole set of BP equations (\ref{eq:BPoriginal}) can be solved by iteration on all the edges of the graph $G$.

At a fixed point of this set of BP equations, we evaluate the marginal probability $\rho_i$ of a vertex $i$ being occupied as
\begin{equation}
  \rho_i \, = \, \frac{ e^{-\beta}}{ e^{-\beta} + \prod\limits_{j\in \partial i}
    q_{j \rightarrow i} }   \; .
\end{equation}
The mean energy density $\rho$ is equal to the mean fraction of occupied vertices,
\begin{equation}
  \rho \, = \, \frac{1}{N} \sum\limits_{i=1}^{N} \rho_i \; .
\end{equation}

The free energy contribution $f_i$ of a vertex $i$ together with all its attached edges is
\begin{equation}
  f_{i} \, = \, - \frac{1}{\beta} \ln \Bigl[ e^{-\beta}
    + \prod\limits_{j \in \partial i} q_{j \rightarrow i} \Bigr] \; ,
\end{equation}
and the free energy contribution $f_{(i, j)}$ of a single edge $(i, j)$ is
\begin{equation}
  f_{(i,j)} \, = \, - \frac{1}{\beta} \ln \Bigl[ q_{i\rightarrow j} +
    q_{j\rightarrow i} \, (1 - q_{i \rightarrow j} ) \Bigr] \; .
\end{equation}
In the free energy summation $\sum_i f_i$ over all the vertices, each edge $(i, j)$ has been counted twice (in $f_i$ and $f_j$). After correcting for this over-counting, the total free energy $F(\beta)$ of the system is evaluated to be
\begin{equation}
  F(\beta ) \, = \, \sum\limits_{i=1}^{N} f_i - \sum\limits_{(j, k)\in G}
  f_{(j, k)} \; .
\end{equation}
The free energy density $f(\beta)$ is simply
\begin{equation}
  f(\beta) \, \equiv \, \frac{ F(\beta) }{N}
  \, = \,  \frac{1}{N} \sum\limits_{i=1}^{N} f_i
  - \frac{1}{N} \sum\limits_{(i, j)\in G} f_{(i, j)} \; .
\end{equation}

At a given energy density $\rho$, the total number of vertex cover configurations is of order $e^{N s}$. Here the entropy density $s$ of the system is evaluated according to
\begin{equation}
  s \, = \, (\rho - f )\, \beta \; .
\end{equation}
The entropy density $s$ is required to be non-negative for this binary state system. This sets a constraint to the inverse temperature $\beta$, namely that $\rho \geq f$ must be satisfied. The value of $\beta$ should not further increase if $\rho$ starts to be less than $f$.

\subsection{Simplified fixed-point equations}

For regular random graphs containing two groups ($A$ and $B$) of vertices with degree parameters $K$ and $K_{\textrm{ba}}$, we assume that the cavity probabilities $q_{j\rightarrow i}$ do not depend on the individual vertex indices but only on the group indices $A$ and $B$. Then,
\begin{equation}
  q_{j\rightarrow i} \, = \,
  \left\{
  \begin{array}{ll}
    q_{\textrm{ab}} \quad \quad \quad & (j\in A , \, 
    i \in B) \; , \\
    q_{\textrm{ba}}  & ( j\in B , \, i \in A) \; , \\
    q_{\textrm{bb}}  & ( j \in B , \, i \in B ) \; .
  \end{array}
  \right.
\end{equation}
According to the BP equation (\ref{eq:BPoriginal}), the values of $q_{\textrm{ab}}$, $q_{\textrm{ba}}$ and $q_{\textrm{bb}}$ are determined through the following three coupled equations
\begin{equation}
  \label{eq:BPrr}
  \begin{aligned}
    q_{\textrm{ab}} & \,= \, \frac{ e^{-\beta} }{ e^{-\beta} + (q_{\textrm{ba}})^{K-1}} \; , \\
    q_{\textrm{ba}} & \, = \, \frac{ e^{-\beta} }{ e^{-\beta} + (q_{\textrm{ab}})^{K_{\textrm{ba}}-1}
      (q_{\textrm{bb}})^{K_{\textrm{bb}}} } \; , \\
    q_{\textrm{bb}} & \, = \, \frac{ e^{-\beta} }{ e^{-\beta} + (q_{\textrm{ab}})^{K_{\textrm{ba}}}
      (q_{\textrm{bb}})^{K_{\textrm{bb}}-1}} \; ,
  \end{aligned}
\end{equation}
where $K_{\textrm{bb}} = K - K_{\textrm{ba}}$. The mean fraction of occupied vertices $\rho$  (the energy density) of the whole system is then
\begin{equation}
  \rho \, = \,
  \frac{ K_{\textrm{ba}} }{ K + K_{\textrm{ba}} } \rho_{\textrm{a}} + \frac{ K }{ K + K_{\textrm{ba}} } \rho_{\textrm{b}}  \; ,
\end{equation}
where $\rho_{\textrm{a}}$ and $\rho_{\textrm{b}}$ are the energy densities of subsystems $A$ and $B$, respectively:
\begin{equation}
  \begin{aligned}
    \rho_{\textrm{a}} & \, = \, \frac{ e^{-\beta}}{e^{-\beta} + (q_{\textrm{ba}})^{K}} \; , \\
    \rho_{\textrm{b}} & \, = \,  \frac{ e^{-\beta}}{e^{-\beta} + (q_{\textrm{ab}})^{K_{\textrm{ba}}}
      (q_{\textrm{bb}})^{K_{\textrm{bb}}}} \; .
  \end{aligned}
\end{equation}
The free energy density $f$ is
\begin{equation}
  \begin{aligned}
    f   \, = \,  & \frac{ K_{\textrm{ba}} }{ K + K_{\textrm{ba}} } f_{\textrm{a}}
    + \frac{ K }{ K + K_{\textrm{ba}} } f_{\textrm{b}}  \\
    &  - \frac{K_{\textrm{ba}}\, K }{ K + K_{\textrm{ba}}} f_{\textrm{ba}} -
    \frac{ K_{\textrm{bb}}\, K}{2 ( K + K_{\textrm{ba}} ) } f_{\textrm{bb}} \; ,
  \end{aligned}
\end{equation}
where the individual free energy expressions are
\begin{equation}
  \begin{aligned}
    f_{\textrm{a}} & \, = \,
    - \frac{1}{\beta} \ln\bigl[ e^{-\beta} + (q_{\textrm{ba}})^K \bigr] \; ,
    \\
    f_{\textrm{b}} & \, = \,
    - \frac{1}{\beta} \ln\bigl[ e^{-\beta} +(q_{\textrm{ab}})^{K_{\textrm{ba}}} \,
      (q_{\textrm{bb}})^{K_{\textrm{bb}}} \bigr] \; , \\
    f_{\textrm{ba}} & \, = \,
    - \frac{1}{\beta} \ln\bigl[ q_{\textrm{ab}} + 
      (1-q_{\textrm{ab}}) \, q_{\textrm{ba}} \bigr] \; , \\
    f_{\textrm{bb}} & \, = \,
    - \frac{1}{\beta} \ln\bigl[ q_{\textrm{bb}} + 
      ( 1 - q_{\textrm{bb}} ) \, q_{\textrm{bb}} \bigr] \; .
  \end{aligned}
\end{equation}

\subsection{The disordered symmetric (DS) fixed point}

The simplified set of BP equations (\ref{eq:BPrr}) has a trivial fixed point, $q_{\textrm{ab}} = q_{\textrm{ba}} = q_{\textrm{bb}} = q$, with $q$ being the root of
\begin{equation}
  \label{eq:qRSrr}
  q \,  = \, \frac{e^{-\beta}}{ e^{-\beta} + q^{K-1}} \; .
\end{equation}
This trivial fixed point corresponds to the disordered symmetric (DS) phase in which every vertex has the same probability $\rho_{D S}$ of being occupied,
\begin{equation}
  \rho_{D S} \, =\, \frac{ e^{-\beta} }{ e^{-\beta} + q^K } \; .
\end{equation}
Microscopic configurations $\vec{\bm{c}}$ in this DS phase contains no information about the planted vertex cover solution. Notice that $\rho_{\textrm{a}} = \rho_{\textrm{b}} = \rho$ in the DS phase. If we define the order parameter $m$ (the magnetization) as
\begin{equation}
  m \equiv \rho_{\textrm{b}} - \rho_{\textrm{a}} \, = \,
  \frac{K + K_{\textrm{ba}}}{K_{\textrm{ba}}} \bigl( \rho_{\textrm{b}} - \rho \bigr) \; ,
  \label{eq:opm}
\end{equation}
then $m=0$ for the DS phase. The DS phase corresponds to the paramagnetic phase of the Ising model.

\subsection{The canonical polarized (CP)  and microcanonical polarized (MP) fixed points}
\label{sec:ggamma}

\begin{figure}
  \centering
  \subfigure[]{
    \includegraphics[angle=270,width=0.49\linewidth]{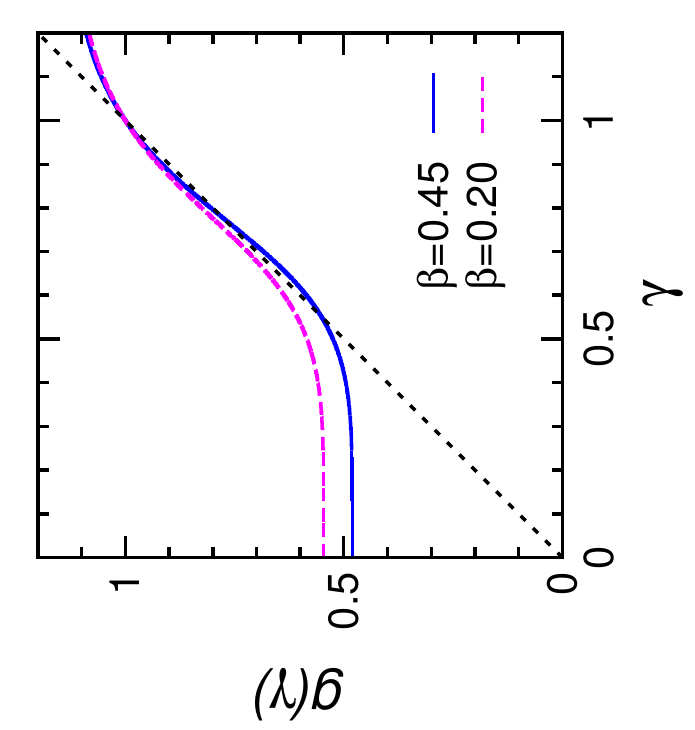}
  }
  \subfigure[]{
    \includegraphics[angle=270,width=0.441\linewidth]{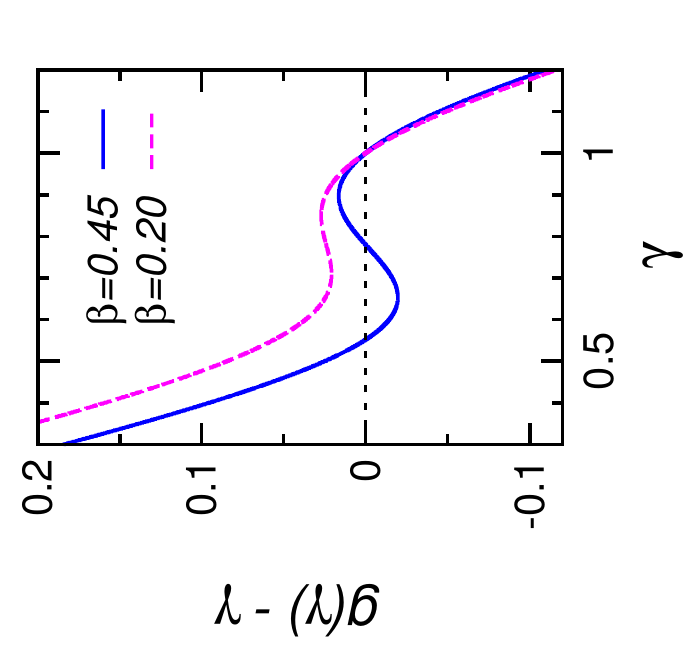}
  }
  \caption{
    \label{fig:ggamma}
    The mapping $g(\gamma)$ as defined by Eq.~(\ref{eq:ggamma}) for $K=10$ and $K_{\textrm{ba}}=6$.  There is always a trivial root $\gamma=1$ for the equation $\gamma = g(\gamma)$. When $\beta > 0.3351$ two additional non-negative roots appear, one of them always being less than unity. The two curves are obtained at $\beta = 0.20$ (dashed) and $\beta=0.45$ (solid). (a) $g(\gamma)$ versus $\gamma$. (b) $g(\gamma) - \gamma$ versus $\gamma$.
  }
\end{figure}

To compute precisely the other fixed points of Eq.~(\ref{eq:BPrr}), we define the ratios $\gamma_{\textrm{ab}}$, $\gamma_{\textrm{ba}}$, and $\gamma_{\textrm{bb}}$ as
\begin{equation}
  \gamma_{\textrm{ab}} \, = \, \frac{q_{\textrm{ab}}}{q} \; , \quad
  \gamma_{\textrm{ba}} \, = \, \frac{q_{\textrm{ba}}}{q} \; , \quad
  \gamma_{\textrm{bb}} \, = \, \frac{q_{\textrm{bb}}}{q} \; .
\end{equation}
The corresponding self-consistent equations are
\begin{subequations}
  \begin{align}
    \gamma_{\textrm{bb}} & \, = \,
    \frac{1}{q + (1 - q) (\gamma_{\textrm{ab}})^{K_{\textrm{ba}}}
      (\gamma_{\textrm{bb}})^{K_{\textrm{bb}}-1} } \; , 
    \label{eq:gammabb}
    \\
    \gamma_{\textrm{ba}} & \, = \,
    \frac{1}{q + (1 - q) (\gamma_{\textrm{ab}})^{K_{\textrm{ba}} - 1}
      (\gamma_{\textrm{bb}})^{K_{\textrm{bb}}} } \; ,
    \label{eq:gammaba}
    \\
    \gamma_{\textrm{ab}} & \, = \,
    \frac{1}{q + (1 - q) (\gamma_{\textrm{ba}})^{K - 1} } \; .
    \label{eq:gammaab} 
  \end{align}
\end{subequations}
Given a value of $\gamma_{\textrm{ab}}$, we can determine the corresponding value of $\gamma_{\textrm{bb}}$ from Eq.~(\ref{eq:gammabb}) and then the corresponding value of $\gamma_{\textrm{ba}}$ through Eq.~(\ref{eq:gammaba}), and afterwords a new value of $\gamma_{\textrm{ab}}$ can be obtained through Eq.~(\ref{eq:gammaab}). In other words, the above three equations form a mapping of $\gamma_{\textrm{ab}}$ to itself. Let us denote this mapping as $g(\gamma_{\textrm{ab}})$. Then the value of $\gamma_{\textrm{ab}}$ is determined by the fixed-point equation
\begin{equation}
  \gamma_{\textrm{ab}} \, = \, g( \gamma_{\textrm{ab}} ) \; .
  \label{eq:ggamma}
\end{equation}
Some example curves of this mapping are shown in Fig.~\ref{fig:ggamma}. The BP fixed points are obtained by finding all the non-negative roots of  Eq.~(\ref{eq:ggamma}).

There is always the trivial solution $\gamma_{\textrm{ab}}^{\textrm{DS}} = 1$ which corresponds to the paramagnetic DS solution. When $\beta$ exceeds certain critical value, two nontrivial solutions $\gamma_{\textrm{ab}}^{\textrm{CP}}$ and $\gamma_{\textrm{ab}}^{\textrm{MP}}$ with $\gamma_{\textrm{ab}}^{\textrm{CP}} < \gamma_{\textrm{ab}}^{\textrm{MP}}$ emerge. The smaller solution $\gamma_{\textrm{ab}}^{\textrm{CP}}$ is less than unity, and it is always locally stable under small perturbations (the slope of $g(\gamma_{\textrm{ab}})$ at $\gamma_{\textrm{ab}}^{\textrm{CP}}$ is less than unity). We refer to this solution the canonical-polarized (CP) fixed point~\cite{Zhou-2019}. It predicts that $\rho_{\textrm{b}}$ is close to unity and $\rho_{\textrm{a}}$ is close to zero, hence a large positive order parameter $m$. This CP phase corresponds to the ferromagnetic phase of the Ising model. It contains vertex cover configurations $\vec{\bm{c}}$ which are highly informative of the planted vertex cover $\Gamma_0$.

The larger solution $\gamma_{\textrm{ab}}^{\textrm{MP}}$ also corresponds to a polarized BP fixed point($\rho_{\textrm{a}} \neq \rho_{\textrm{b}}$). Following Ref.~\cite{Zhou-2019}, we refer to it as the microcanonical-polarized (MP) fixed point. The nature of the thermodynamic phase described by this MP fixed point will be discussed in detail in the next two sections. We note that at a fixed inverse temperature $\beta$, the free energy density $f_{M P}$ of the MP fixed point is higher than the value $f_{C P}$ of the CP phase. If $\gamma_{\textrm{ab}}^{\textrm{MP}} < 1$ then it is a locally unstable solution of Eq.~(\ref{eq:ggamma}) as the slope of $g(\gamma_{\textrm{ab}})$ exceeds unity at $\gamma_{\textrm{ab}}^{\textrm{MP}}$, and its magnetization $m$ is a small positive value. If $\gamma_{\textrm{ab}}^{\textrm{MP}} > 1$, then this MP fixed point is a locally stable solution of Eq.~(\ref{eq:ggamma}) as the slope of $g(\gamma_{\textrm{ab}})$ is less than unity at $\gamma_{\textrm{ab}}^{\textrm{MP}}$ and it is the polarized metastable phase complementary to the CP phase, with negative magnetization $m$.

We now describe some representative results predicted by the mean field theory. To be concrete we will first consider three graph ensembles with fixed vertex degree $K=10$. We choose three different values for the inter-group degree ($K_{\textrm{ba}} = 7, 6, 5$), which lead to qualitatively different properties. We also consider another graph ensemble $K=5$, $K_{\textrm{ba}}=4$ which is qualitatively similar to the graph ensemble of $K=10$ and $K_{\textrm{ba}}=6$.

\section{The graph ensemble with degree parameters $K = 10$ and $K_{\textrm{ba}}=5$}
\label{sec:10_5_5}

\subsection{Equilibrium canonical phase transition}

\begin{figure}
  \centering
  \hspace*{-0.15cm}
  \subfigure[]{
    \label{fig:K10v5v5:a}
    \includegraphics[angle=270,width=0.49\linewidth]{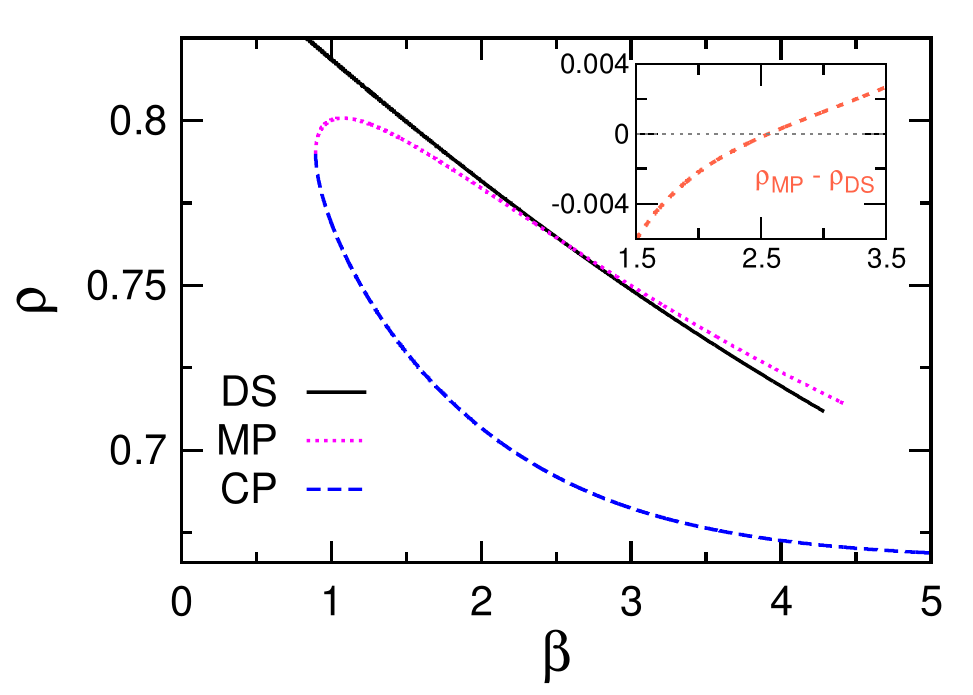}
  }   \hspace*{-0.4cm}
  \subfigure[]{
    \label{fig:K10v5v5:b}
    \includegraphics[angle=270,width=0.49\linewidth]{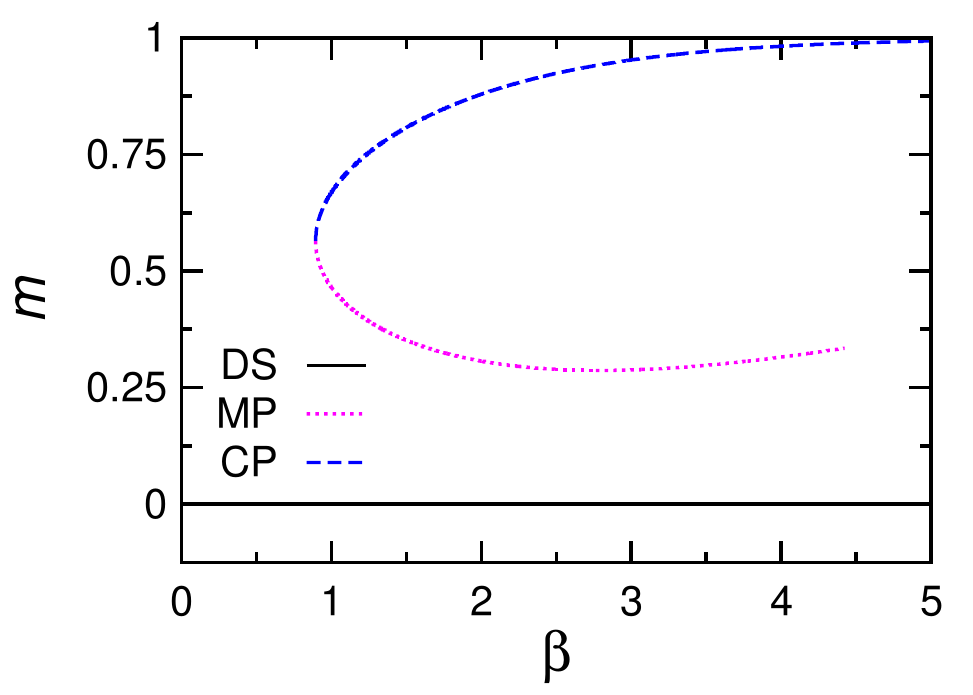}
  }
  \\
  \hspace*{-0.15cm}
  \subfigure[]{
    \label{fig:K10v5v5:c}
    \includegraphics[angle=270,width=0.49\linewidth]{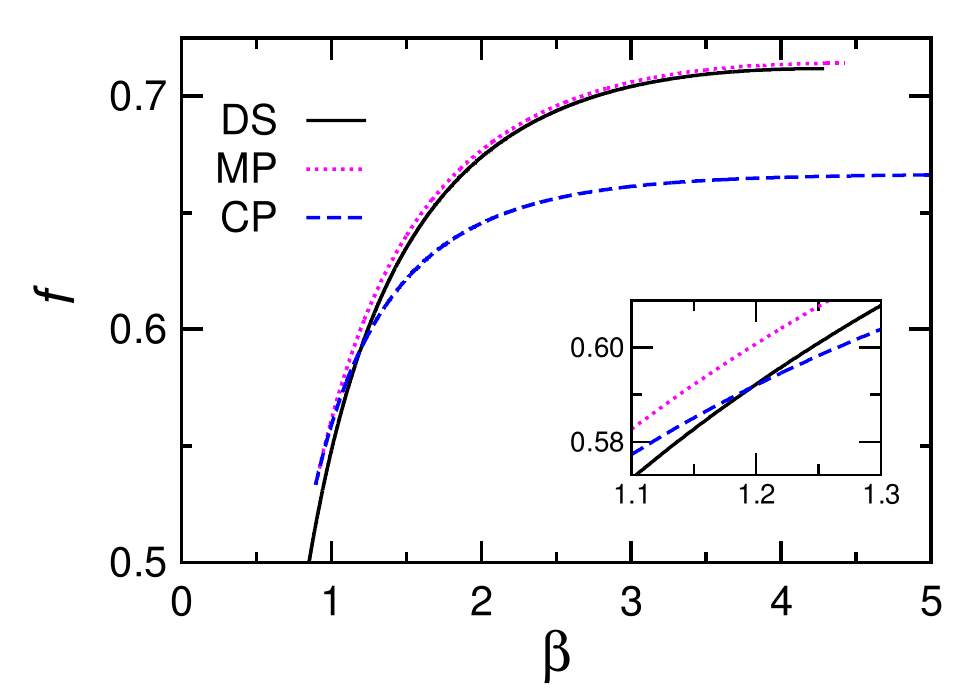}
  }   \hspace*{-0.4cm}
  \subfigure[]{
    \label{fig:K10v5v5:e}
    \includegraphics[angle=270,width=0.49\linewidth]{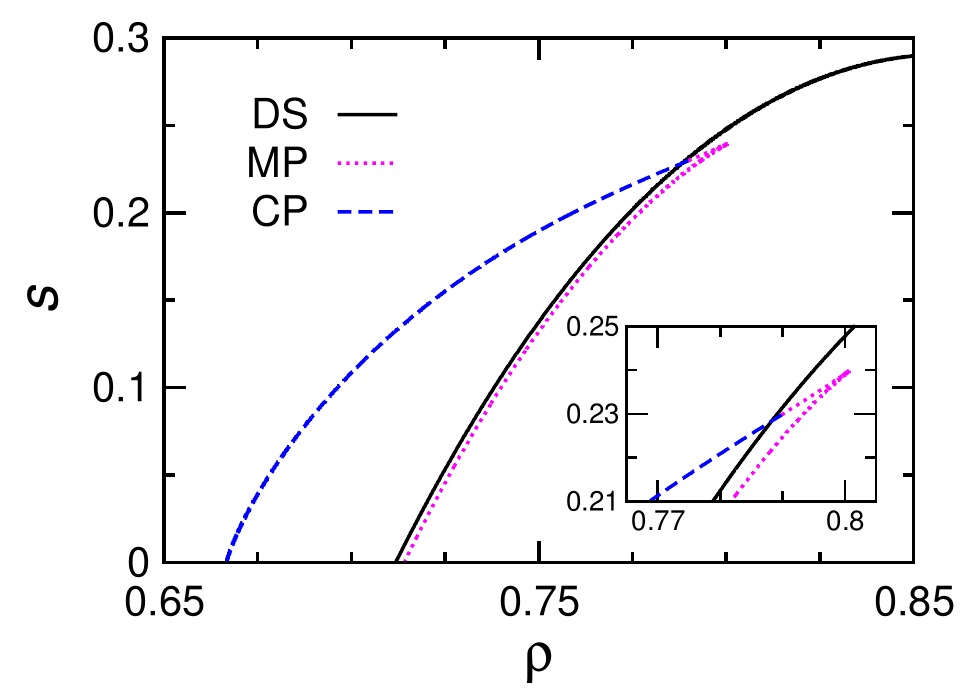}
  }
  \caption{
    Theoretical results on the planted regular random graph ensemble with degree parameters $K=10$, $K_{\textrm{ba}} = 5$. Results corresponding to the disordered symmetric (DS), microcanonical polarized (MP), and canonical polarized (CP) fixed points are distinguished by different line and color types. (a) Energy density $\rho$ versus inverse temperature $\beta$. Inset shows the difference of energy density $\Delta \rho = \rho_{\textrm{MP}}-\rho_{\textrm{DS}}$ between the MP and DS phases. (b) Magnetization $m$ versus $\beta$. (c) Free energy density $f$ versus $\beta$. Inset is a magnified view of the region $\beta \in (1.1, 1.3)$. (d) Entropy density $s$ versus $\rho$. Inset is a magnified view of the region $\rho \in (0.765, 0.805)$.
  }
  \label{fig:K10v5v5}
\end{figure}

The mean field theoretical results obtained for the ensemble of $K=10$ and $K_{\textrm{ba}} = K_{\textrm{bb}} = 5$ are summarized in Fig.~\ref{fig:K10v5v5}. When the inverse temperature $\beta$ exceeds $0.8957$, the ferromagnetic CP phase starts to emerge in the configuration space of vertex covers with a large positive value of magnetization $m_{\textrm{CP}} > 0.5$. The mean energy density $\rho_{\textrm{CP}}$ of the CP phase is much lower than that of the paramagnetic DS phase. The free energy density $f_{\textrm{CP}}$ of the CP phase starts to be lower than its counterpart $f_{\textrm{DS}}$ of the DS phase as $\beta$ exceeds the value $\beta_{F} = 1.1958$ (Fig.~\ref{fig:K10v5v5:c}, inset), indicating that an equilibrium ferromagnetic phase transition occurs between the DS and CP phases at this critical point $\beta_{F}$. This equilibrium phase transition is a discontinuous one, with a discontinuity in the mean energy density $\rho$.  At the equilibrium ferromagnetic transition (DS $\rightarrow$ CP) the order parameter $m$ jumps from zero to a large positive value.

\subsection{Free energy barrier}

The MP fixed point appears together with the CP fixed point at $\beta = 0.8957$ and its mean energy density $\rho_{\textrm{MP}}$ is larger than $\rho_{\textrm{CP}}$. Interestingly, we find that $\rho_{\textrm{MP}}$ of the MP fixed point starts to exceed the energy density $\rho_{\textrm{DS}}$ of the DS phase as $\beta$ goes beyond $2.5580$ (Fig.~\ref{fig:K10v5v5:a}, inset). This turns out to have algorithmic implication as we will shortly discuss. The magnetization of the MP fixed point is positive and it lies between the corresponding values for the paramagnetic DS and the ferromagnetic CP phases (Fig.~\ref{fig:K10v5v5:b}), and the free energy density $f_{\textrm{MP}}$ of the MP fixed point is higher than those of the DS and CP phases in the whole range of $\beta$ (Fig.~\ref{fig:K10v5v5:c}). These theoretical results suggest that the MP phase (the configuration subspace described by the MP fixed point) is the watershed separating the ferromagnetic CP and the paramagnetic DS phases.

\begin{figure}
  \centering
  \hspace*{-0.15cm}
  \subfigure[]{
    \label{fig:K10v5v5:dbf}
    \includegraphics[angle=270,width=0.49\linewidth]{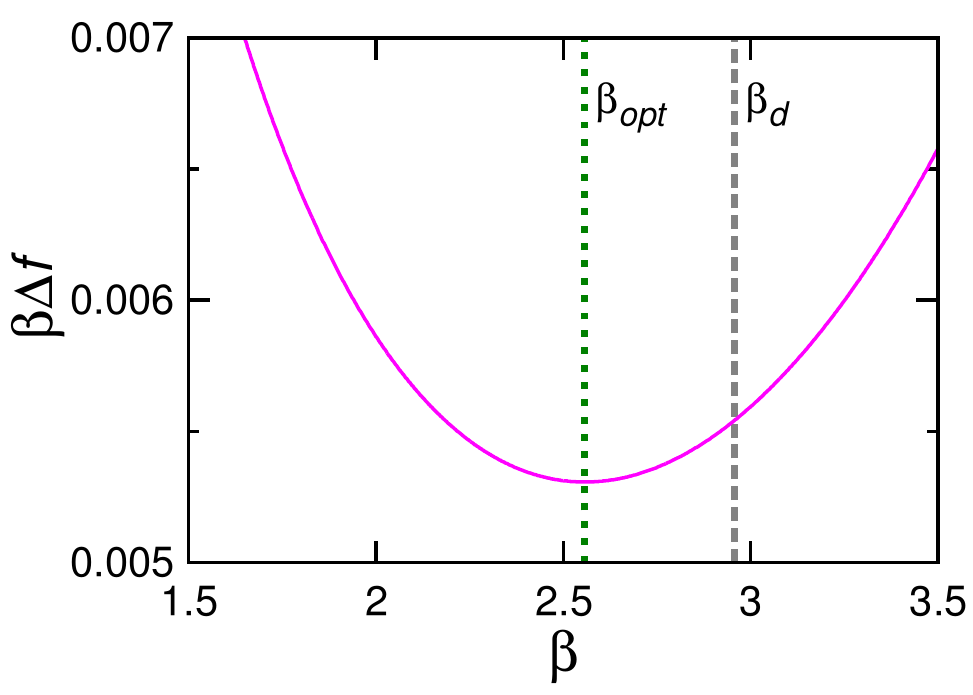}
  }   \hspace*{-0.4cm}
  \subfigure[]{
    \label{fig:K10v5v5:ds}
    \includegraphics[angle=270,width=0.49\linewidth]{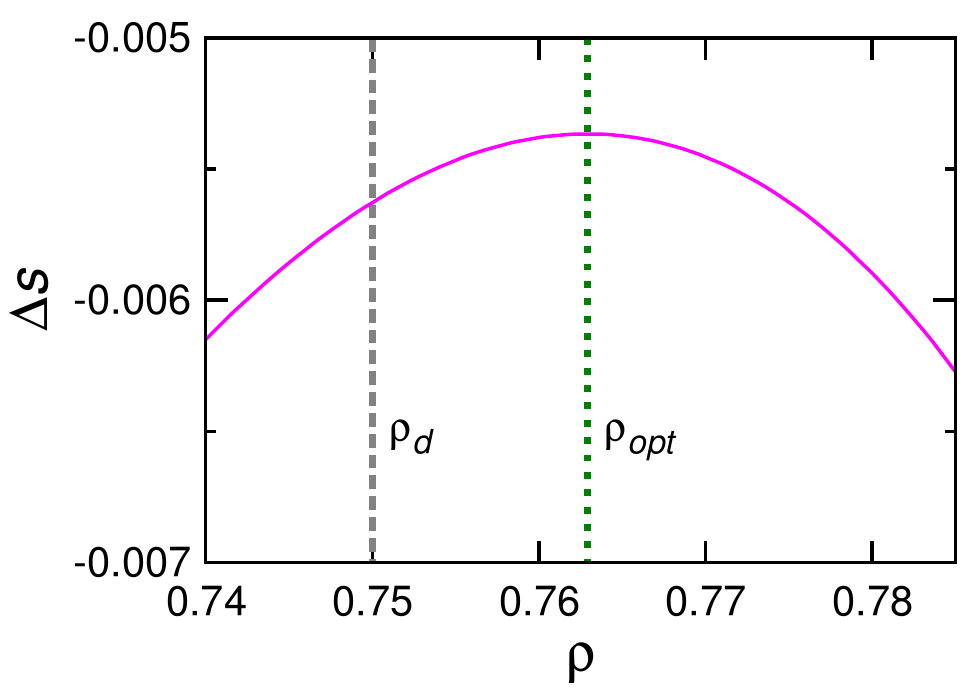}
  }
  \caption{
    The rescaled free energy density difference $\beta \Delta f$ (a) and entropy density difference $\Delta s$ (b) between the intermediate MP phase and the paramagnetic DS phase, for the planted regular random graph ensemble with degree parameters $K=10$, $K_{\textrm{ba}}=5$. Vertical dashed lines mark the spin glass dynamic phase transition point ($\beta_d$ and $\rho_d$), vertical dotted lines mark the optimal escaping point ($\beta_{opt}$ and $\rho_{opt}$).
  }
  \label{fig:K10v5v5diff}
\end{figure}

Let us denote the free energy density difference, mean energy density difference, and entropy density difference between the MP and DS phases as $\Delta f$, $\Delta \rho$ and $\Delta s$ respectively,
\begin{subequations}
  \begin{align}
    \Delta f  & \, \equiv \, f_{\textrm{MP}} - f_{\textrm{DS}} \; ,
    \label{eq:df} \\
    \Delta \rho & \, \equiv \, \rho_{\textrm{MP}} - \rho_{\textrm{DS}} \; ,
    \label{eq:dr} \\
    \Delta s & \, \equiv \, s_{\textrm{MP}} - s_{\textrm{DS}} \; .
  \end{align}
  \label{eq:dfdrds}
\end{subequations}
The total free energy difference $N \Delta f$ is the minimum free energy barrier to be overcome to transition from the DS phase to the CP phase. Notice that $\Delta f$ is always positive for the ensemble of $K = 10$, $K_{\textrm{ba}} = 5$. In the thermodynamic limit $N\rightarrow \infty$, the probability of overcoming such an extensive free energy barrier is exponentially small ($\approx e^{-N \beta \Delta f}$). We plot the coefficient $\beta \Delta f$ as a function of $\beta$ in Fig.~\ref{fig:K10v5v5:dbf}, which has a global minimum value at the inverse temperature value $\beta_{opt} = 2.5580$. The theoretical result therefore predicts that $\beta_{opt}$ is the optimal inverse temperature point for traversing the MP free energy barrier.

The optimal transition inverse temperature $\beta_{opt}$ is exactly the inverse temperature at which the mean energy densities of the DS and MP phases become equal (Fig.~\ref{fig:K10v5v5:a}, inset). To prove this, we notice that
\begin{equation}
  \frac{\textrm{d} (\beta \Delta f)}{\textrm{d} \beta}  \, = \, 
  \Delta \rho \; .
  \label{eq:barrierslope}
\end{equation}
The minimum of $\beta \Delta f$ is therefore reached when $\Delta \rho = 0$ (if $\Delta \rho$ itself is an increasing function of $\beta$). At this particular value $\beta_{opt}$ of inverse temperature, the free energy barrier is contributed purely by entropic effect ($\Delta s < 0$ and $\Delta \rho = 0$). We have checked that $\beta_{opt}$ is lower than the dynamical spin glass phase transition critical inverse temperature $\beta_{d} = 2.9560$ of the paramagnetic DS phase (see Sec.~\ref{sec:lsa} and Appendix~\ref{app:dsgt})~\cite{Zhang-Zeng-Zhou-2009}.

\subsection{Equilibrium microcanonical phase transition}

The entropy density functions $s_{\textrm{DS}}(\rho)$ and $s_{\textrm{CP}}(\rho)$ of the DS and CP phases are shown in Fig.~\ref{fig:K10v5v5:e}, and we find that $s_{\textrm{CP}}(\rho)$ starts to be greater than $s_{\textrm{DS}}(\rho)$ when the mean energy density $\rho$ decreases below the critical value $0.7883$. It means that there is an equilibrium microcanonical phase transition between the paramagnetic DS and the ferromagnetic CP phases at this energy density. It is a discontinuous phase transition, similar to what occurs for the Potts model~\cite{Zhou-2019}. At the transition point, the inverse temperature of the CP phase is lower than that of the DS phase, so the phase transition is associated with a discontinuity in the inverse temperature $\beta$. To escaping the DS phase at fixed mean energy density $\rho$, the system has to overcome the watershed as located by the lower branch of the MP curve (Fig.~\ref{fig:K10v5v5:e}, inset), and the entropy barrier is $\Delta s = - 0.0066$ at the phase transition point. 

\subsection{Entropy barrier}

The microcanonical transition from the paramagnetic DS phase to the ferromagnetic CP phase can also be carried out at other values of mean energy density $\rho$. We plot in Fig.~\ref{fig:K10v5v5:ds} the entropy barrier $\Delta s$ as a function of $\rho$. We see that $\Delta s(\rho)$ is negative and its maximum value $-0.0053$ is located at the optimal value $\rho_{opt} = 0.7629$. In the thermodynamic limit the probability of overcoming the MP entropy barrier will be exponentially small and be of the order $\exp( N \Delta s )$.   Therefore the optimal strategy to carry out the microcanonical transition is to fix the energy density to the optimal value $\rho_{opt}$. We find that this value is larger than the critical value of $\rho_d = 0.75$ at which the paramagnetic DS phase experiences a spin glass dynamic phase transition (Sec.~\ref{sec:lsa} and Appendix~\ref{app:dsgt}). This means that the barrier escaping dynamics at $\rho_{opt}$ will not be complicated by the possibility of spin glass traps.

\subsection{Computer simulation results}
\label{subsec:MC}

We carry out Monte Carlo computer simulations to verify the predictions of the preceding subsections. Our simulation method is very similar to the one employed in Refs.~\cite{Zhou-2019,Zhou-Liao-2021}. Here we only briefly describe the simulation protocol, and put the technical details in Appendix~\ref{app:mcmc}.

To check the existence of an equilibrium discontinuous phase transition between the paramagnetic DS phase and the ferromagnetic CP phase in the canonical ensemble, we use inverse temperature $\beta$ as the control parameter and let it decrease slowly between consecutive evolution epochs. The initial value is set to be $\beta=2.0$ and the initial microscopic configuration is set to be $c_i = 1$ for all the vertices $i$ of group $B$ and $c_j = 0$ for all the vertices $j$ of group $A$ (the planted vertex cover). At later inverse temperature values, the initial microscopic configuration is the last sampled configuration of the previous inverse temperature. To ensure that the system is equilibrated at each value of $\beta$, we first run the Monte Carlo dynamics for a time at least $10^6$ sweeps before the start of actual configuration sampling (one sweep is $N$ consecutive updating trials of the microscopic configuration). 

To check the existence of an equilibrium discontinuous phase transition between the paramagnetic DS phase and the ferromagnetic CP phase (or the polarized MP phase) in the microcanonical ensemble, we use the energy density $\rho$ as the control parameter and let it slowly increase during the simulation process (the initial microscopic configuration is again the planted vertex cover).  At later $\rho$ values, the initial microscopic configuration is the last sampled configuration of the previous $\rho$. We also fully equilibrate the system at each value of $\rho$ before the configuration sampling process.

\begin{figure}
  \centering
  \hspace*{-0.15cm}
  \subfigure[]{
    \includegraphics[angle=270,width=0.49\linewidth]{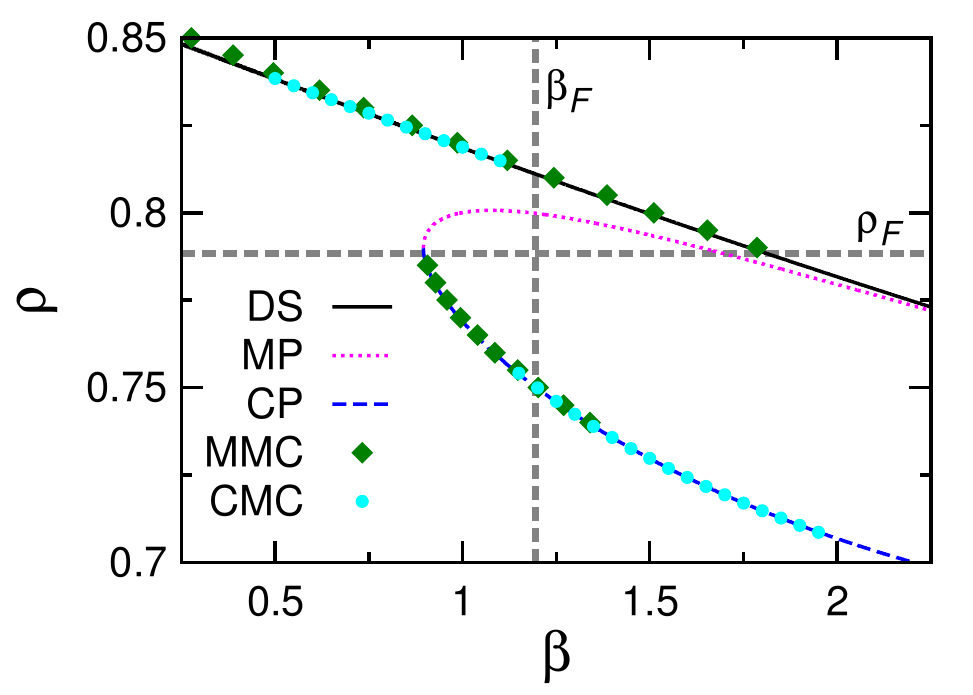}
    \label{fig:MCK1055BvsR}
  } \hspace*{-0.4cm}
  \subfigure[]{
    \includegraphics[angle=270,width=0.49\linewidth]{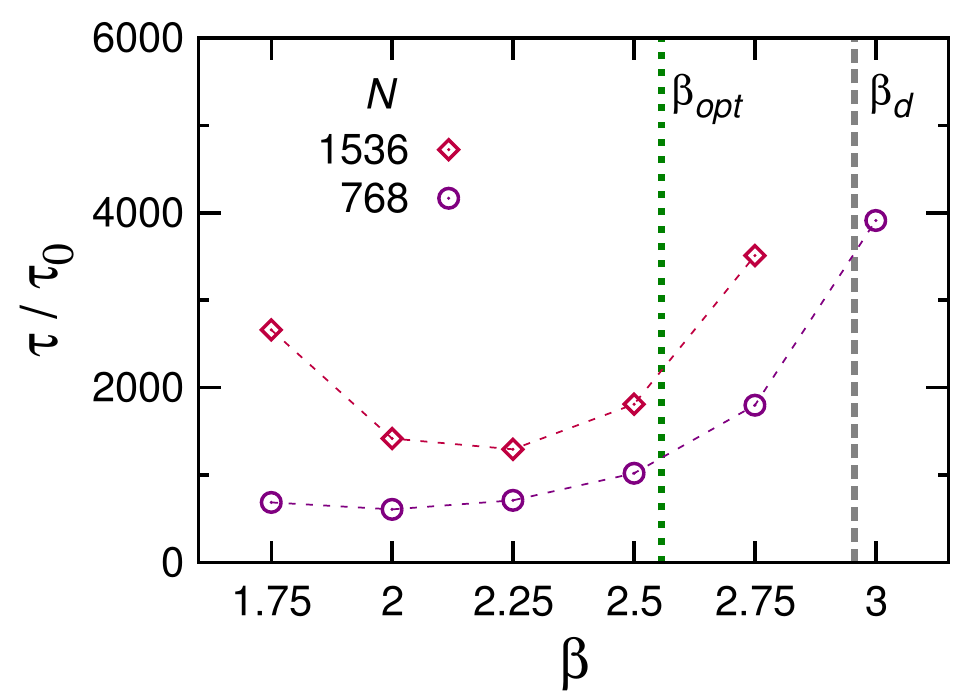}
    \label{fig:ESTK1055}
  }
  \caption{
    Simulation results obtained on single planted regular random graph instances with degree parameters $K=10$ and $K_{B A} = 5$. (a) Mean energy density $\rho$ at fixed inverse temperature $\beta$ (canonical Monte Carlo, circle) and mean inverse temperature $\beta$ at fixed energy density $\rho$ (microcanonical Monte Carlo, diamond), for a single graph containing $N=2100$ vertices. The curves are theoretical predictions as in Fig.~\ref{fig:K10v5v5:a}, and the vertical and horizontal dashed lines indicate the equilibrium phase transition points.  (b) Median waiting time $\tau$ needed to make the transition from the paramagnetic DS phase to the ferromagnetic CP phase at inverse temperature $\beta$, on a single graph of size $N$, whose value is $768$ (circle) and $1536$ (diamond). The escaping time $\tau$ is rescaled by $\tau_0 = e^{0.0053\, N}$. The vertical dotted line marks the predicted optimal inverse temperature $\beta_{opt} = 2.5580$, the vertical dashed line locates the spin glass transition inverse temperature $\beta_d = 2.9560$.
  }
  \label{fig:MCresult}
\end{figure}

Figure~\ref{fig:MCK1055BvsR} summarizes the numerical results obtained on a single graph instance of size $N=2100$. In agreement with the theoretical predictions, we observe a sudden jump of the mean energy density $\rho$ at inverse temperature $\beta \approx 1.20$ in the canonical evolution trajectory, and a sudden jump of the inverse temperature $\beta$ at energy density $\rho \approx 0.79$ in the microcanonical evolution trajectory. We have also performed computer simulations on some other graph ensembles (data not shown), and the discontinuous nature of the equilibrium ferromagnetic phase transitions are always confirmed.

To confirm the existence of an optimal escaping inverse temperature $\beta_{opt}$ as predicted in Fig.~\ref{fig:K10v5v5:dbf}, we perform escaping dynamics on single planted random graph instances of different sizes $N$. For a given graph instance, we first sample $60$ independent equilibrium vertex cover configurations of the paramagnetic DS phase at fixed inverse temperature $\beta$. Each of these initial configurations has been fully equilibrated  within the paramagnetic DS phase, and we checked that it has no net magnetization ($m=0$) and its energy density $\rho$ is equal to the predicted theoretical value. To reach equilibrium within the paramagnetic DS phase, we employ two types of elementary trials under the vertex-cover constraints and the condition of detailed balance: flipping ($c_i: 0 \leftrightarrow 1$ for vertex $i$) and swapping ($c_i: 0 \rightarrow 1$ for an empty vertex $i$ followed by $c_j: 1\rightarrow 0$ for an occupied vertex $j$). The swapping trial enables long-distance exchange of vertex states and it has been highly helpful in structural glass studies~\cite{Berthier-etal-2018}. The frequencies of swapping and flipping trials are $0.9$ and $0.1$, respectively, in our implementation. The paramagnetic DS phase contains microscopic configurations whose fractions of occupied vertices in the two groups are almost equal ($\rho_{\textrm{a}} \approx \rho_{\textrm{b}}$). To ensure that the whole evolution trajectory stays in the paramagnetic DS phase, we reject all the inter-group swaps which will enlarge the difference between $\rho_{\textrm{a}}$ and $\rho_{\textrm{b}}$.

The fractions of occupied vertices in group $A$ and group $B$ are equal ($\rho_{\textrm{a}} = \rho_{\textrm{b}}$) in each of these $60$ initial equilibrium DS configurations. Then staring from each of these $60$ initial configurations, we perform canonical Monte Carlo evolution by single vertex flipping trials (without employing the swapping trials) at the same value of $\beta$ without imposing the constraint of $\rho_{\textrm{a}} = \rho_{\textrm{b}}$, and run $1000$ independent stochastic evolution trajectories. One unit time corresponds to $N$ consecutive single-vertex flipping attempts. For each trajectory, we record the earliest evolution time $\tau$ at which the energy density $\rho$ drops below the threshold value $\rho_{\theta} = 0.95\, \rho_{\textrm{CP}} + 0.05\, \rho_{\textrm{DS}}$, where $\rho_{\textrm{CP}}$ and $\rho_{\textrm{DS}}$ are the mean energy densities of the CP and DS phases at $\beta$, respectively.

The median value of $\tau$ among these $60000$ trajectories is shown in Fig.~\ref{fig:ESTK1055}. In this figure, we rescaled the first-passage time by $\tau_0 = e^{0.0053 N}$ in accordance with the minimum value of $\beta \Delta f = 0.0053$ (Fig.~(\ref{fig:K10v5v5:dbf}). The simulation results clearly confirm that there exists an optimal inverse temperature to escape the paramagnetic DS phase. However, the empirically observed optimal inverse temperature is lower than the theoretically predicted optimal value of $2.5580$.

The discrepancy may be due to kinetic effects which are not included in the mean field theory. An empty vertex temporarily blocks all its nearest neighbors from changing to the empty state. Such kinetic blocking effects become more and more stronger as $\beta$ increases, simply because more vertices are empty at higher $\beta$ values. We expect that the kinetic rate of configuration updates will decrease faster than $e^{-\beta}$ as $\beta$ increases, and this may contribute a large prefactor $a_k$ to the scaling behavior $\tau \simeq a_k e^{N \beta \Delta f}$.  The kinetic effects on $a_k$ should be less significant in comparison with the thermodynamic effect $e^{N \beta \Delta f}$ as system size $N$ increases.  Indeed, as $N$ increases from $768$ to $1536$, the empirically determined optimal $\beta$ increases from $\beta_{opt} \approx 2$ to $\beta_{opt} \approx 2.25$ (Fig.~\ref{fig:K10v5v5:dbf}). We expect that the empirical $\beta_{opt}$ will be much more closer to the theoretically predicted value as $N$ becomes sufficiently large.

The nonmonotonic temperature dependence of the first-passage time to escaping the paramagnetic phase has some apparent similarity with the the phenomenon of nonmonotonic temperature dependence of crystallization observed in glass-forming liquids~\cite{Schmelzer-etal-2016}. Another apparently related issue in supercooled liquids is the Mpemba effect~\cite{Jeng-2006,Zhang-Hou-2022} which states that certain liquid may freeze faster starting from a high temperature than starting from a low temperature. These similarities deserve to be further investigated.

As a brief summary of the whole section, we regard the planted regular random graph ensemble with $K=10$, $K_{\textrm{ba}}=5$ as an example system with the free energy landscape of type Fig.~\ref{fig:LSt}.

\section{The graph ensemble with degree parameters $K=10$ and $K_{\textrm{ba}} = 7$}
\label{sec:10_7_3}

\begin{figure}[b]
  \centering
  \hspace*{-0.15cm}
  \subfigure[]{
    \label{fig:K10v3v7:BE}
    \includegraphics[angle=270,width=0.49\linewidth]{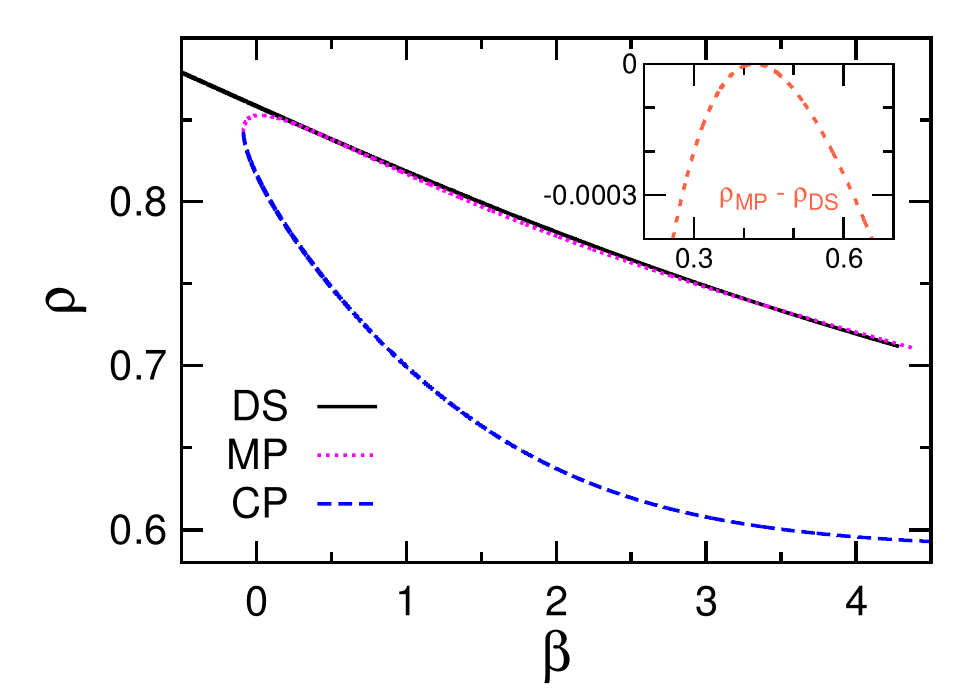}
  } \hspace*{-0.4cm}
  \subfigure[]{
    \label{fig:K10v3v7:BM}
    \includegraphics[angle=270,width=0.49\linewidth]{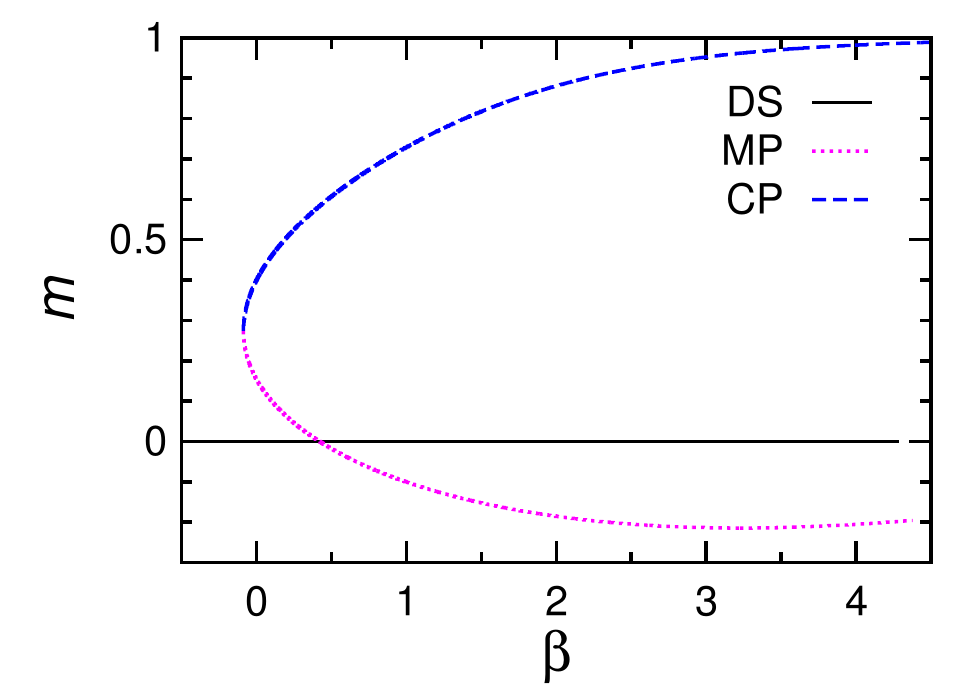}
  }
  \\
  \hspace*{-0.15cm}
  \subfigure[]{
    \label{fig:K10v3v7:BF}
    \includegraphics[angle=270,width=0.49\linewidth]{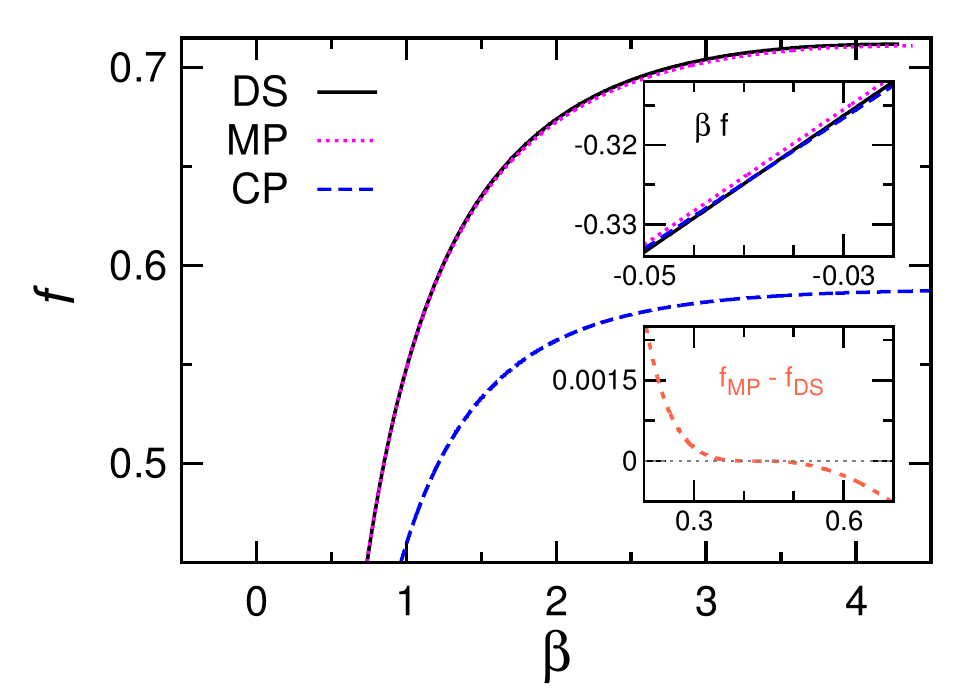}
  } \hspace*{-0.4cm}
  \subfigure[]{
    \label{fig:K10v3v7:ES}
    \includegraphics[angle=270,width=0.49\linewidth]{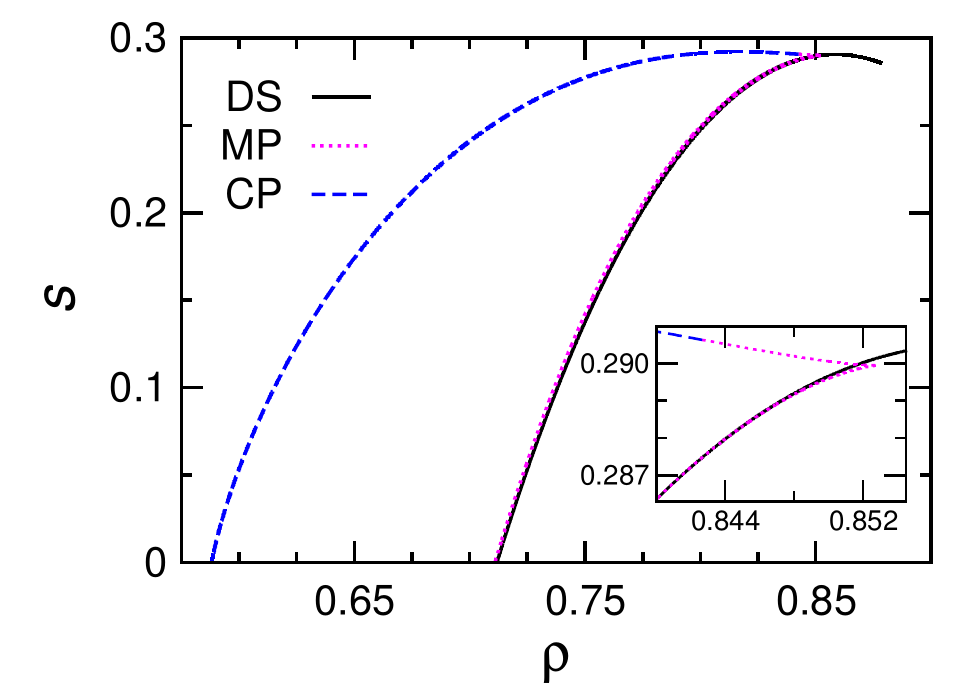}
  }
  \caption{
    \label{fig:K10v3v7}
    Theoretical results on the planted graph ensemble of $K=10$, $K_{\textrm{ba}} = 7$. (a) Energy density $\rho$. Notice that $\rho_{\textrm{MP}} \leq \rho_{\textrm{DS}}$. (b) Magnetization $m$. (c) Free energy density $f$. The top inset plots $\beta f$ at the vicinity of $\beta = -0.04$. The bottom inset shows the free energy density difference $f_{\textrm{MP}} - f_{\textrm{DS}}$ at the vicinity of $\beta = 0.42$. (d) Entropy density $s$. The region of $\rho \approx 0.847$ is magnified.
  }
\end{figure}

We now focus on another representative planted random graph ensemble whose uniform degree is $K=10$ and whose inter-group degree is $K_{\textrm{ba}} = 7$ (each vertex in group $B$ is connected to seven vertices of group $A$ and to three vertices of group $B$). We summarize the mean field theoretical results for this graph ensemble in Fig.~\ref{fig:K10v3v7}.

First, we notice that the ferromagnetic CP phase starts to emerge at the negative inverse temperature $\beta = -0.0854$ with a high energy density $\rho = 0.8427$ and a moderate positive magnetization $m=0.2741$. The paramagnetic DS  phase is characterized by zero magnetization ($m=0$), and its energy density is higher than that of the CP phase at any inverse temperature $\beta$. Concerning the free energy densities of these two phases, we find that $\beta f_{\textrm{CP}}$ starts to be lower than $\beta f_{\textrm{DS}}$ at $\beta = \beta_{F}= -0.0382$ (Fig.~\ref{fig:K10v3v7:BF}, top inset), and when $\beta > 0$ the free energy density $f_{\textrm{CP}}$ of the CP phase is always lower than the corresponding $f_{\textrm{DS}}$ of the DS phase (Fig.~\ref{fig:K10v3v7:BF}). These results mean that there exists a discontinuous equilibrium phase transition between the DS and CP phases at the critical inverse temperature $\beta_{F}$ (the ferromagnetic phase transition). For all inverse temperatures $\beta > \beta_{F}$ (which include the whole region of positive $\beta$) the CP phase is the unique true equilibrium thermodynamic state.

We now examine the MP fixed point of the mean field theory. There is a critical inverse temperature value $\beta_{\textrm{pf}} = 0.4213$ at which the MP and DS fixed points coincide with each other. Comparing with the result in Fig.~\ref{fig:K10v5v5:a} for the ensemble of $K_{\textrm{ba}}=5$, we find that at $K_{\textrm{ba}}=7$ the mean energy density $\rho_{\textrm{MP}}$ of the MP phase is lower than the corresponding $\rho_{\textrm{DS}}$ of the DS phase at both sides of this critical $\beta_{\textrm{pf}}$ value (Fig.~\ref{fig:K10v3v7:BE}, inset). For $\beta < \beta_{\textrm{pf}}$, the MP phase has intermediate positive magnetization $m$ between those of the DS and CP phases (Fig.~\ref{fig:K10v3v7:BM}) and its free energy density is higher than those of the DS and CP phases (Fig.~\ref{fig:K10v3v7:BF}, bottom inset). These results indicate that the MP phase is the free-energy barrier (watershed) which separates the DS phase from the CP phase. This free energy density barrier to escape the DS phase (Eq.~(\ref{eq:df})) drops to zero at $\beta = \beta_{\textrm{pf}}$. Local stability analysis reveals that at $\beta = \beta_{\textrm{pf}}$ the paramagnetic DS phase becomes locally unstable towards the ferromagnetic CP phase (see Sec.~\ref{sec:lsa}).  We will argue shortly that, in the thermodynamic limit $N\rightarrow \infty$, this inverse temperature $\beta_{opt} = \beta_{\textrm{pf}}$ is the optimal choice to achieve the quickest transition from the DS phase to the CP phase by a stochastic local search dynamics. 

When $\beta$ exceeds the critical value $\beta_{\textrm{pf}}$, the MP phase changes to have negative magnetization ($m < 0$) and its free energy density $f_{\textrm{MP}}$ becomes lower than that of the DS phase (Fig.~\ref{fig:K10v3v7:BF}, bottom inset). These results indicate that the MP fixed point at $\beta > \beta_{\textrm{pf}}$ describes a metastable anti-ferromagnetic phase in which vertices of group $A$ are more densely occupied than vertices of group $B$. The paramagnetic DS phase then serves as the watershed in the free energy landscape separating this anti-ferromagnetic MP phase and the ferromagnetic CP phase. Notice that the free energy density barrier height ($f_{\textrm{DS}}-f_{\textrm{MP}}$) increases from zero as $\beta$ increases from $\beta_{\textrm{pf}}$ (Fig.~\ref{fig:K10v3v7:BF}, bottom inset), meaning that the chance of escaping the anti-ferromagnetic MP phase is exponentially rare ($\sim e^{-N \beta (f_{\textrm{DS}}-f_{\textrm{MP}})}$) in the thermodynamic limit. As the paramagnetic DS phase is unstable for the whole inverse temperature range $\beta > \beta_{\textrm{pf}}$, if starting from an initial configuration of the DS phase and let it evolve at fixed $\beta > \beta_{\textrm{pf}}$, with some positive probability it will slide to the anti-ferromagnetic MP phase and then getting trapped. So it is not advisable to set the search $\beta$ to be higher than $\beta_{\textrm{pf}}$. In the thermodynamic limit $N\rightarrow \infty$, the unique optimal $\beta$ value for accessing the planted vertex cover solution is to set the inverse temperature to be $\beta_{\textrm{pf}}$, at which there is no risk of being trapped by the paramagnetic DS phase or by the anti-ferromagnetic MP phase and there is no extensive free energy barrier ($\Delta f = 0$). 

The entropy densities $s$ of the DS, MP and CP fixed points of the BP equation are plotted in Fig.~\ref{fig:K10v3v7:ES} as functions of energy density $\rho$. We notice that, when $\rho \in [0.8427, 0.8529)$ the MP entropy density function has two branches (Fig.~\ref{fig:K10v3v7:ES}, inset). The upper entropy branch is convex in $\rho$, meaning that the corresponding phase is only stable in the microcanonical ensemble with fixed energy density. This entropy branch crosses with the entropy density function of the paramagnetic DS phase at the critical energy density $\rho = 0.8517$, indicating that there is a discontinuous microcanonical equilibrium phase transition, with a discontinuity of inverse temperature from $\beta = 0.1651$ (DS) to $\beta = -0.0374$ (MP). There is an extensive entropic barrier at this microcanonical phase transition, whose magnitude could be quantified by the entropy density difference ($= -1.2\times 10^{-4}$) between the entropy density of the lower MP branch and that of the DS phase. In the thermodynamic limit this microcanonical transition is impossible to occur. However, as the energy density decreases from $0.8517$, the entropy barrier shrinks and it finally vanishes at $\rho = 0.8413$, indicating that the optimal energy density to escaping the DS phase is $\rho = \rho_{opt} = 0.8413$. Notice that this optimal density $\rho_{opt}$ is exactly corresponding to the optimal inverse temperature $\beta_{opt} = \beta_{\textrm{pf}} = 0.4213$.

Local stability analysis (see Sec.~\ref{sec:lsa}) reveals that, at $\beta = \beta_{\textrm{pf}}$ the paramagnetic DS phase is still locally stable towards disordered spin glass phases. The critical inverse temperature of local instability towards spin glass phases is  $\beta_{\textrm{pg}}=2.9560$.
  
As a brief summary of this section, we regard the planted regular random graph ensemble with $K=10$, $K_{\textrm{ba}}=7$ as an example of systems with the free energy landscape of type Fig.~\ref{fig:LUns}, in which there is an exchange of free energy valley with free energy saddle point at the inverse temperature $\beta_{\textrm{pf}}$, with $\beta_{\textrm{pf}}$ being the point of local instability of the paramagnetic DS phase.

\section{Special planted regular random graph ensembles}
\label{sec:special}

\subsection{The case of $K = 10$ and $K_{\textrm{ba}}=6$}
\label{sec:10_6_4}

\begin{figure}
  \centering
  \hspace*{-0.15cm}
  \subfigure[]{
    \label{fig:K10v4v6:BE}
    \includegraphics[angle=270,width=0.49\linewidth]{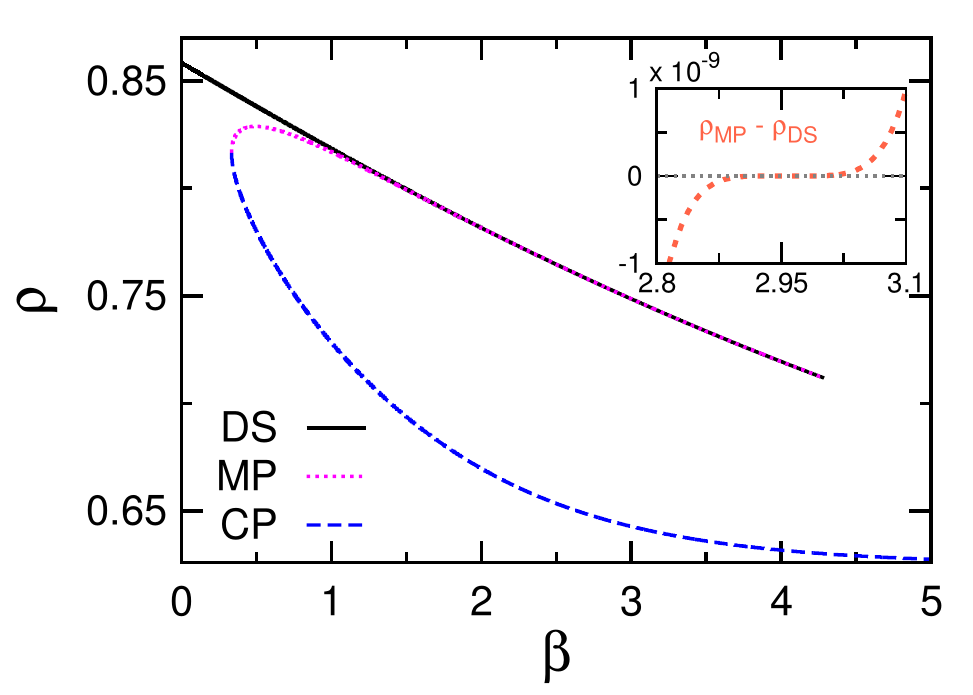}
  } \hspace*{-0.4cm}
  \subfigure[]{
    \label{fig:K10v4v6:BM}
    \includegraphics[angle=270,width=0.49\linewidth]{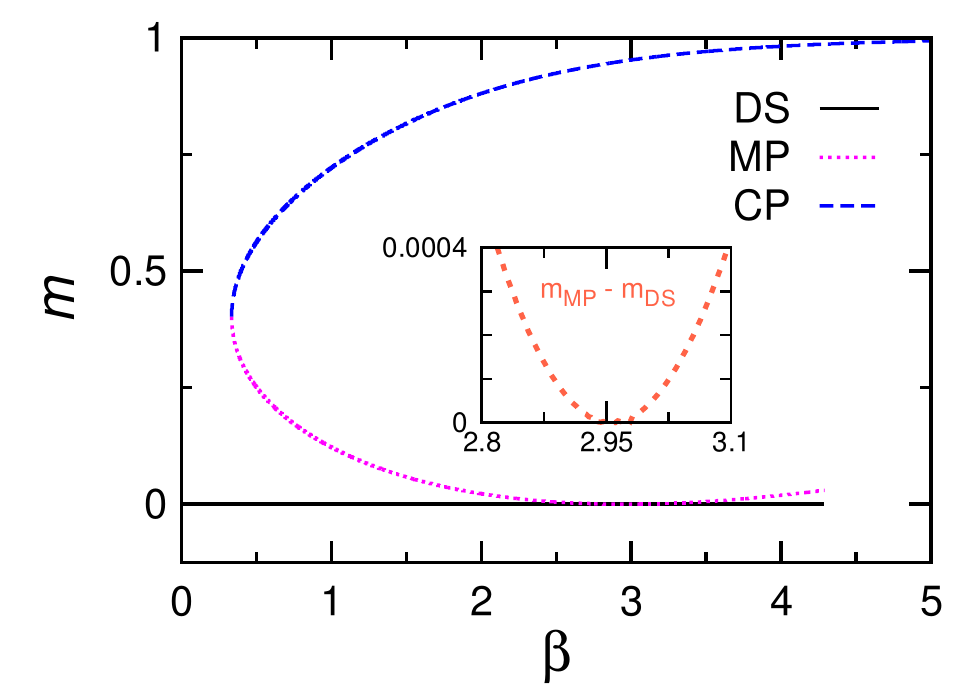}
  }
  \\
  \hspace*{-0.15cm}
  \subfigure[]{
    \label{fig:K10v4v6:BF}
    \includegraphics[angle=270,width=0.49\linewidth]{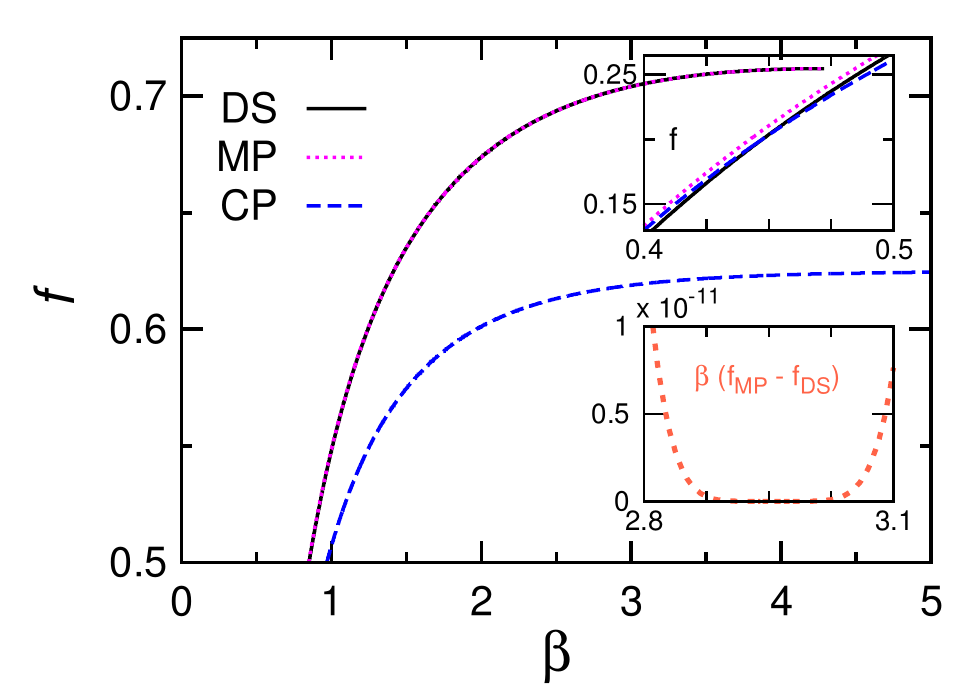}
 } \hspace*{-0.4cm}
 \subfigure[]{
    \label{fig:K10v4v6:ES}
    \includegraphics[angle=270,width=0.49\linewidth]{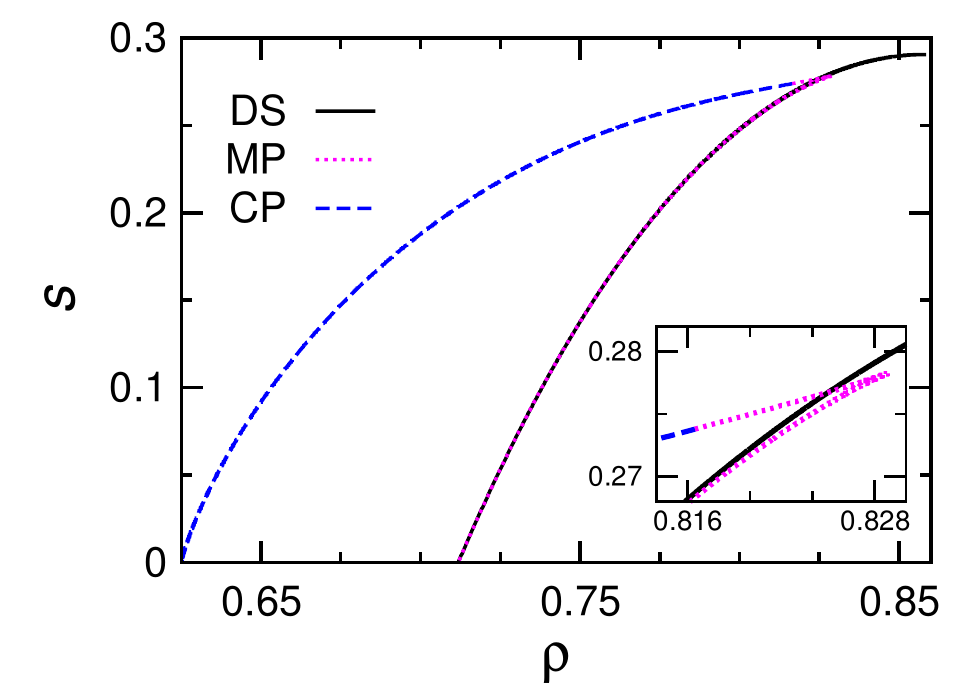}
 }
  \caption{
    \label{fig:K10v4v6}
    Theoretical results on the special planted regular random graph ensemble with degree parameters $K=10$, $K_{\textrm{ba}} = 6$. (a) Energy density $\rho$. The inset shows the difference $\rho_{\textrm{MP}} - \rho_{\textrm{DS}}$ at $\beta \approx 2.95$. (b) Magnetization $m$. The inset shows the difference $m_{\textrm{MP}}-m_{\textrm{DS}}$ at $\beta \approx 2.95$. (c) Free energy density $f$. The region of $\beta \approx 0.45$ is shown in the top inset, and the bottom inset shows the value of $\beta( f_{\textrm{MP}} - f_{\textrm{DS}})$ at the vicinity of $\beta = 2.95$. (d) Entropy density $s$. The inset highlight the entropy crossing at $\rho = 0.8250$.
  }
\end{figure}

We now focus on the special planted regular random graph ensemble of degree $K=10$ which lies between the two ensembles discussed in the preceding two sections. In this special case, each vertex in group $B$ is connected to $K_{\textrm{ba}} = 6$ vertices of group $A$ and to $K_{\textrm{bb}} = 4$ other vertices of group $B$. Figure~\ref{fig:K10v4v6} collects the main theoretical results obtained on this graph ensemble. In the following we will emphasize the distinctive features of this new ensemble in comparison with the ensembles of $K_{\textrm{ba}} = 5$ and $K_{\textrm{ba}} = 7$.

The polarized CP and MP fixed points start to emerge at inverse temperature $\beta = 0.3351$. As $\beta$ increases, we find that the energy density difference $\Delta \rho$ between the MP and DS phases of this $K_{\textrm{ba}} = 6$ ensemble changes from being negative to being positive at inverse temperature $\beta = 2.9560$ and the rescaled free energy density difference $\beta (f_{\textrm{MP}} - f_{\textrm{DS}})$ is minimized at this point (insets of Fig.~\ref{fig:K10v4v6:BE} and ~\ref{fig:K10v4v6:BF}), indicating the optimal inverse temperature to access the planted ferromagnetic CP phase from the paramagnetic DS phase is $\beta=2.9560$. This same feature of nonmonotonic $\beta \Delta f$ has also been observed for the ensemble of $K_{\textrm{ba}} = 5$, but here the minimum value is exactly zero. 

At $\beta = 2.9560$ the MP and DS fixed points merge with each other, leading to vanishing magnetization $m_{\textrm{MP}}$ of the MP phase and vanishing free energy difference $(f_{\textrm{MP}} - f_{\textrm{DS}})$. This merging behavior is identical to that of the $K_{\textrm{ba}} = 7$ ensemble, and it indicates that the DS phase at this inverse temperature is in a critical state (it turns out that this paramagnetic phase starts to be locally unstable, see Sec.~\ref{sec:lsa}). But different from the $K_{\textrm{ba}} = 7$ ensemble, as $\beta$ exceeds $2.9560$ the magnetization $m_{\textrm{MP}}$ does not change to be negative but change back to be positive (Fig.~\ref{fig:K10v4v6:BM}), and the rescaled free energy difference $\beta \Delta f$ becomes positive again (Fig.~\ref{fig:K10v4v6:BF}, bottom inset). Therefore the $K_{\textrm{ba}} = 6$ ensemble does not have a stable anti-ferromagnetic phase, but instead the paramagnetic DS phase recovers its local stable with respect to the MP fixed point at inverse temperatures $\beta > 2.9560$.

The entropy density profiles of the DS, CP and MP phases are compared in Fig.~\ref{fig:K10v4v6:ES}. From these results and the data shown in Fig.~\ref{fig:K10v4v6:BF}, we conclude that the equilibrium canonical phase transition from the paramagnetic DS phase to the ferromagnetic CP phase occurs at the critical inverse temperature $\beta_{F} = 0.4485$ (Fig.~\ref{fig:K10v4v6:BF}, top inset), at which the mean energy density drops from $\rho_{\textrm{DS}} = 0.8401$ to $\rho_{\textrm{CP}} = 0.7878$. Of course, this equilibrium phase transition can not be observed in the $N\rightarrow \infty$ limit due to the extensive free energy barrier caused by the intermediate MP phase (Fig.~\ref{fig:K10v4v6:BF}, bottom inset). The entropy profile of the MP phase has two branches in the range of $\rho \in [0.8165, 0.8288]$ and the upper branch intersects with the DS phase at $\rho = 0.8250$ (Fig.~\ref{fig:K10v4v6:ES}, inset), meaning that there exists an equilibrium microcanonical phase transition from the paramagnetic DS phase to the polarized MP phase at this critical energy density. This is a discontinuous phase transition with a drop of the inverse temperature from $\beta_{\textrm{DS}} = 0.8335$ to $\beta_{\textrm{MP}} = 0.3666$. This equilibrium microcanonical phase transition also can not be observed in the thermodynamic limit due to the huge entropy barrier caused by the lower-branch MP watershed~\cite{Zhou-2019}.  

\subsection{The general degree condition}

The free energy landscape of the graph ensemble with $K=10$, $K_{\textrm{ba}}=6$ can be schematically illustrated by Fig.~\ref{fig:LMr}. For this ensemble, we find that the maximum slope of the function $g(\gamma_{\textrm{ab}})$ (defined in Sec.~\ref{sec:ggamma}) at $\gamma_{\textrm{ab}} = 1$ is exactly equal to unity. This property guarantees that the MP fixed-point $\gamma_{\textrm{ab}}^{\textrm{MP}}$ of Eq.~(\ref{eq:ggamma}) will never go beyond unity but be on the same side of $\gamma_{\textrm{ab}}^{\textrm{CP}}$ (which is less than unity).  After some elementary derivations, we figure out that an infinite number of other planted regular random graph ensembles also have the same property.  The degree parameters of these special graph ensembles are governed by
\begin{equation}
  K \, = \, (\ell + 1 )^2 + 1 \; , \quad K_{\textrm{ba}} \, = \, 2 (\ell + 1 ) \; ,
  \label{eq:Kscaling} 
\end{equation}
with integer $\ell = 1, 2, \ldots$, and they satisfy the equality
\begin{equation}
  K_{\textrm{ba}}^2 \, = \,  4 (K - 1) \; .
  \label{eq240924a}
\end{equation}
We list the first few such ensembles in Table~\ref{tab:specialRR}.

\begin{table}[b]
  \caption{
    \label{tab:specialRR}
    Examples of special planted regular random graph ensembles whose degrees are specified by Eq.~(\ref{eq:Kscaling}). The critical inverse temperature $\beta_{\textrm{pf}}$ is determined by Eq.~(\ref{eq:betab}), the corresponding energy density is $\rho_{\textrm{pf}} = 1/(1+(K-1)^{-1/2})$. The minimum fraction of occupied vertices $\rho_{0} = K/(K + K_{\textrm{ba}})$.
  }
  \centering
  \vskip 0.2cm
  \begin{tabular}{lllll}
    \hline \hline
    $K$ & $K_{\textrm{ba}}$\quad \quad   & $\beta_{\textrm{pf}}$ & $\rho_{\textrm{pf}}$ & $\rho_{0}$  \\
    \hline
    $5$     & $4$  & $2.7726$  \quad \quad & $0.6667$ \quad \quad & $0.5556$\\
    $10$       & $6$  & $2.9560$ &       $0.75$ & $0.625$ \\
    $17$       & $8$        & $3.5043$ &       $0.8$ & $0.68$\\
    $26$       & $10$       & $4.1923$ &       $0.8333$ & $0.7222$  \\   
    $37$       & $12$       & $4.9541$ &       $0.8571$ & $0.7551$ \\
    $50$       & $14$       & $5.7616$ &       $0.875$  & $0.7813$\\
    $65$       & $16$       & $6.6001$ &       $0.8889$ & $0.8025$\\
    $82$       & $18$       & $7.4610$ &       $0.9$ & $0.82$\\
    $101$\quad \quad & $20$ & $8.3388$ &       $0.9091$ &  $0.8347$ \\
    \hline \hline
  \end{tabular}
\end{table}

When we perform local stability analysis (see the next section~\ref{sec:lsa}) on the paramagnetic DS phase, we find that Eq.~(\ref{eq240924a}) is exactly the condition to guarantee $\beta_{\textrm{pf}} = \beta_{\textrm{pg}}$. The critical inverse temperature $\beta_{\textrm{pf}}$ (respectively, $\beta_{\textrm{pg}}$) for the local instability of the paramagnetic DS phase towards the ferromagnetic CP phase (respectively, disordered spin glass phases) can be exactly computed as
\begin{equation}
  \beta_{\textrm{pf}} = \beta_{\textrm{pg}} \, = \, 
  (\ell + 1)^2 \ln ( 1 + 1/ \ell ) - \ln \ell \; .
  \label{eq:betab}
\end{equation}
The numerical value is $\beta_{\textrm{pf}} = 2.9560$ for $K=10$, $K_{\textrm{ba}}=6$ of the preceding subsection. The intermediate MP fixed point keeps to be locally unstable at $\beta > \beta_{\textrm{pf}}$. The free energy landscape of type Fig.~\ref{fig:LMr} has therefore only a single stable ordered CP phase (besides the various disordered glass phases), at $\beta > \beta_{\textrm{pf}}$. Both the DS and MP fixed points are only saddle points of this free energy landscape. 

Because of the existence of extensive free energy barrier ($\beta \Delta f > 0$) at $\beta < \beta_{\textrm{pf}}$ on the one hand and the existence of extremely many disordered spin glass phases at $\beta > \beta_{\textrm{pf}}$ ($=\beta_{\textrm{pg}}$) on the other hand, a natural prediction of the mean field theory is that the inverse temperature $\beta = \beta_{\textrm{pf}}$ will be the only point at which a stochastic local search dynamics, starting from an initial microscopic configuration of the paramagnetic DS phase, can hope to reach the planted ferromagnetic CP phase in polynomial time (in the thermodynamic limit $N\rightarrow \infty$). We will now check this prediction by numerical simulations. We will work on the most sparse regular random graph ensemble of Table~\ref{tab:specialRR}, namely the one with $K=5$ and $K_{\textrm{ba}}=4$, which makes it possible to simulate larger systems at relatively low computation cost.

\subsection{The most sparse case of $K=5$ and $K_{\textrm{ba}}=4$}
\label{sec:5_4_1}

\begin{figure}
  \centering
  \hspace*{-0.15cm}
  \subfigure[]{
    \label{fig:K5v4v1:rvsb}
    \includegraphics[angle=270,width=0.49\linewidth]{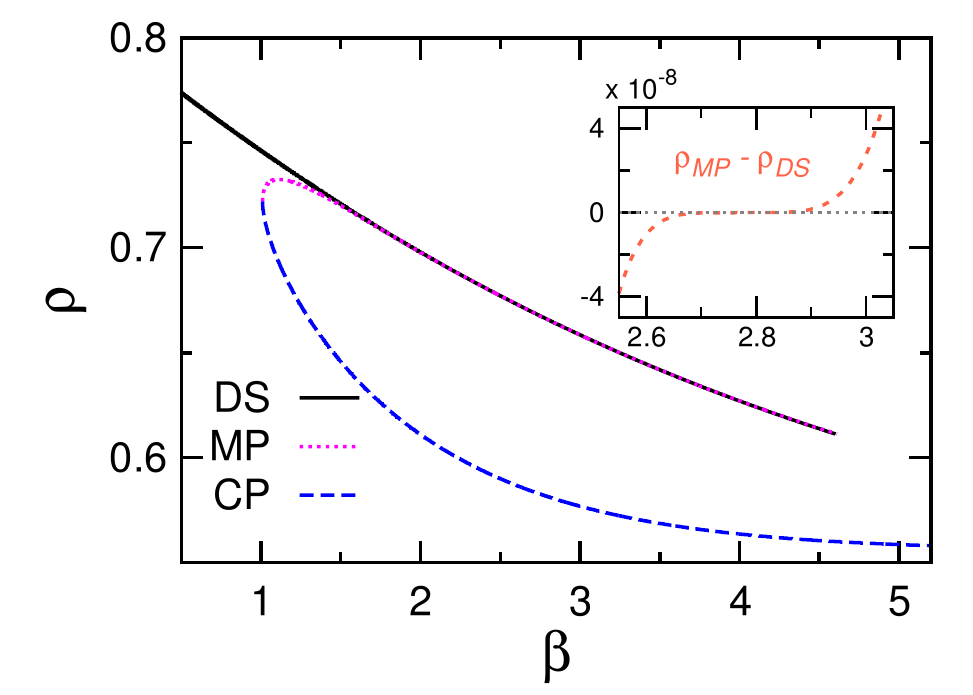}
  } \hspace*{-0.4cm}
  \subfigure[]{
    \label{fig:K5v4v1:mvsb}
    \includegraphics[angle=270,width=0.49\linewidth]{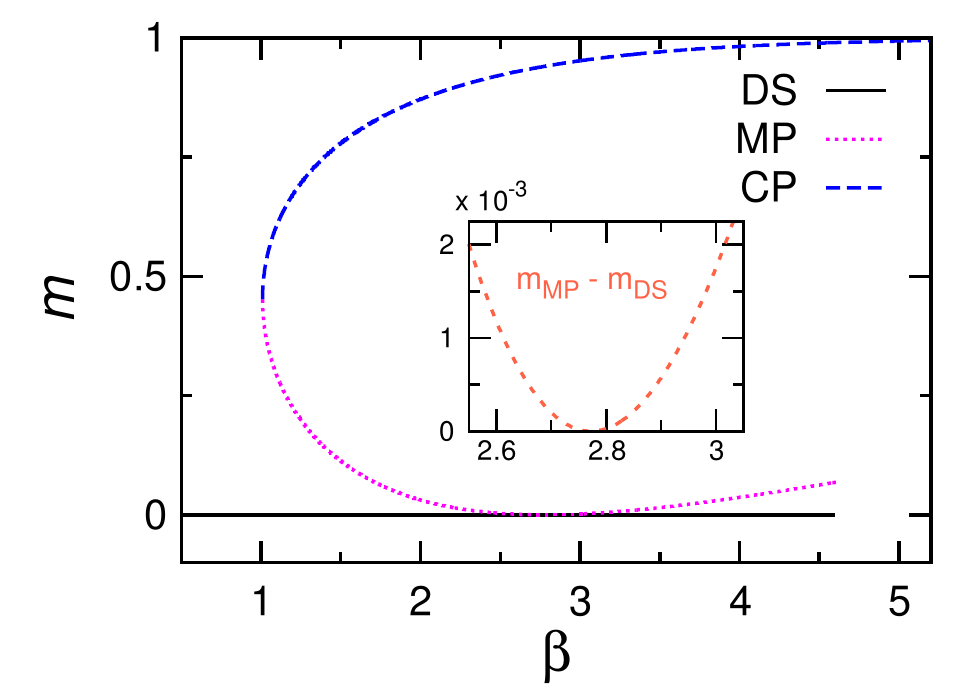}
  }
  \\
  \hspace*{-0.15cm}
  \subfigure[]{
    \label{fig:K5v4v1:fvsb}
    \includegraphics[angle=270,width=0.49\linewidth]{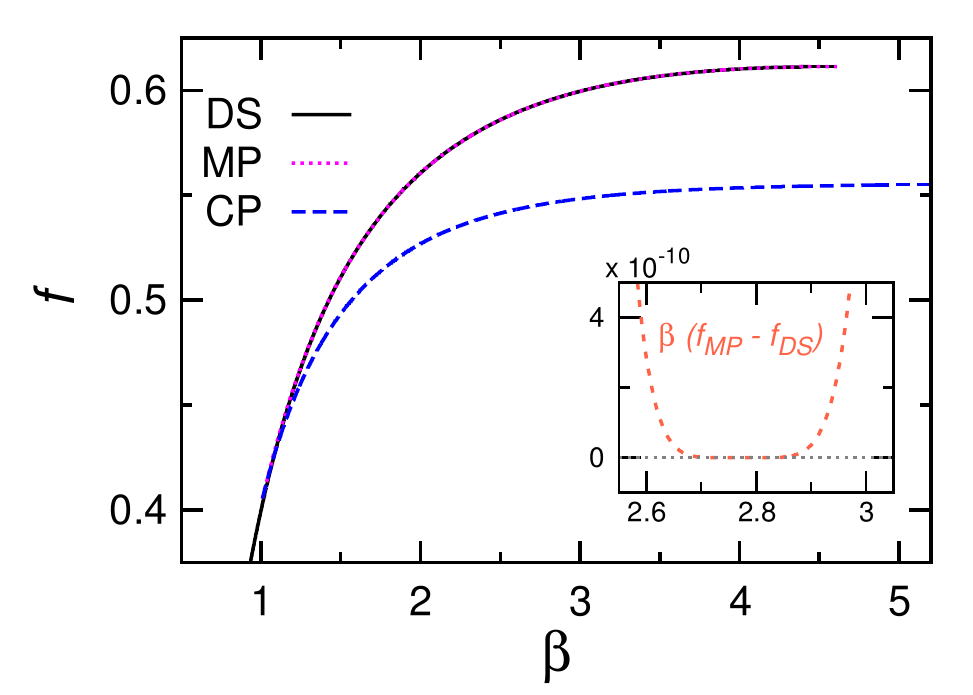}
  }   \hspace*{-0.4cm}
  \subfigure[]{
    \label{fig:K5v4v1:svse}
    \includegraphics[angle=270,width=0.49\linewidth]{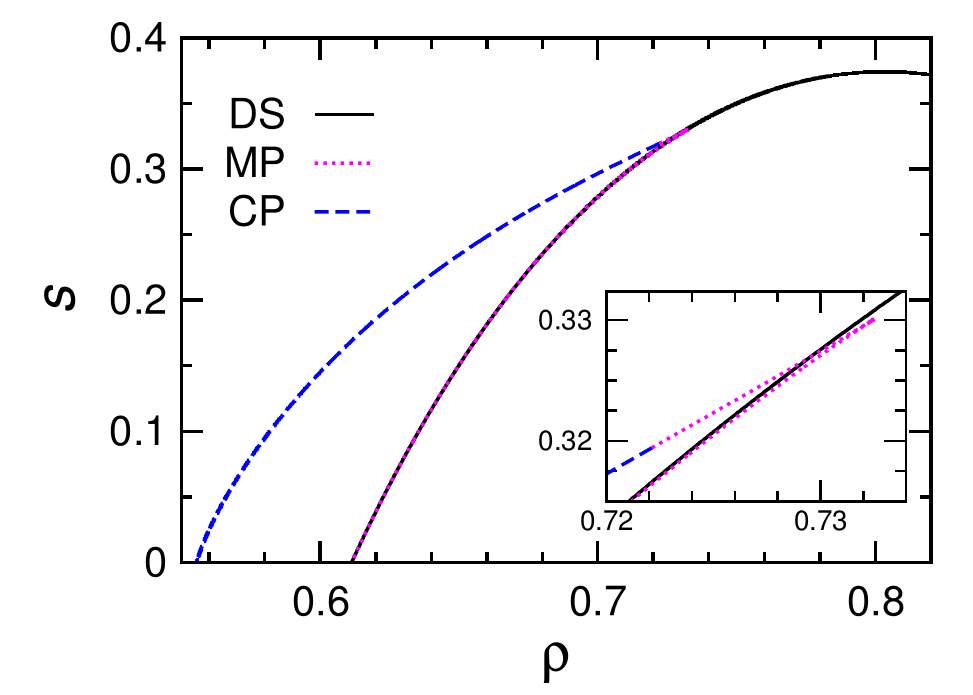}
  }
  \\
  \subfigure[]{
    \includegraphics[angle=270,width=0.49\linewidth]{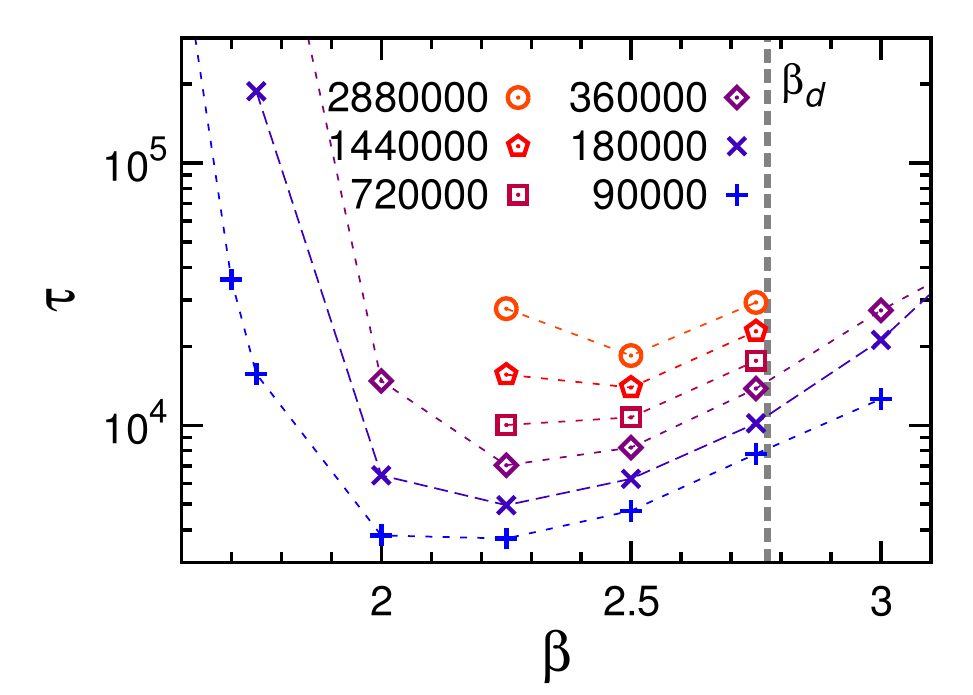}
    \label{fig:estimek5v4}
  }
  \caption{
    Theoretical and simulation results on the special planted regular random graph ensemble with degree parameters $K=5$, $K_{\textrm{ba}} = 4$. (a)--(d): Same as Fig.~\ref{fig:K10v4v6}, for energy density $\rho$, magnetization $m$, free energy density $f$, and entropy density $s$. (e)  Median first-passage waiting time $\tau$ to observe a DS-to-CP transition at inverse temperature $\beta$ by a local dynamical process, among $60000$ independent evolution trajectories on a single graph instance with size $N$ ranging from $90\,000$ to $2\,880\,000$. Dotted vertical line indicates the critical inverse temperatures $\beta_{d}  = \beta_{\textrm{pg}} = \beta_{\textrm{pf}} = 2.7726$.
  }
  \label{fig:K5v4v1}
\end{figure}

The predicted energy density $\rho$, magnetization $m$, free energy density $f$, and entropy density $s$ of the special graph ensemble with $K = 5$ and $K_{\textrm{ba}}=4$ are reported in Fig.~\ref{fig:K5v4v1} for the DS, MP and CP phases, which are qualitatively similar to the results of Fig.~\ref{fig:K10v4v6} for the ensemble of $K=10$ and $K_{\textrm{ba}}=6$. The free energy density of the ferromagnetic CP phase becomes lower than that of the paramagnetic DS phase at the critical inverse temperature $\beta_F = 1.0800$, leading to a discontinuous equilibrium ferromagnetic phase transition. There is also a discontinuous microcanonical phase transition between the paramagnetic DS and partially polarized MP phases when their entropy densities become equal at the critical energy density $\rho = 0.7295$ (Fig.~\ref{fig:K5v4v1:svse}, inset). Both these equilibrium phase transitions have been verified by computer simulations on systems of relatively small sizes. When system size $N$ becomes large, due to the extensive free energy or entropy barrier, the waiting times needed to observe such equilibrium transitions are exponentially large and these transitions are therefore dynamically blocked.

The paramagnetic DS phase becomes unstable with respect to the ferromagnetic CP phase and the disordered spin glass phases at the critical inverse temperature $\beta_{\textrm{pf}} = \beta_{\textrm{pg}} = 2.7726$. The rescaled MP-DS free energy barrier $\beta \Delta f$ reaches the minimum value of zero at this inverse temperature (Fig.~\ref{fig:K5v4v1:fvsb}, inset), indicating that the optimal inverse temperature for a stochastic local search dynamics to escape from the paramagnetic DS phase to the ferromagnetic CP phase is located at $\beta = 2.7726$.

To check this theoretical prediction, we perform Monte Carlo simulations at fixed values of $\beta$ and estimate the first-passage time for the energy density to reach the equilibrium value of the ferromagnetic CP phase, starting from a set of $60$ initial microscopic configurations which are located in the paramagnetic DS phase. Each of these initial configurations has been fully equilibrated within the paramagnetic DS phase by adopting the same protocol as described in Sec.~\ref{subsec:MC}. To estimate the median first-passage time from such an initial equilibrium DS configuration at inverse temperature $\beta$ to a final configuration which is located in the ferromagnetic CP phase, we consider a sequence of single-spin flipping trials under the condition of detailed balance. One time unit corresponds to $N$ consecutive single-spin flipping trials. Again, an escaping process is considered as successfully achieved when the energy density $\rho$ of the microscopic configuration has reached below the threshold value $\rho_\theta = 0.95 \rho_{\textrm{CP}} + 0.05 \rho_{\textrm{DS}}$ for the first time, where $\rho_{\textrm{CP}}$ and $\rho_{\textrm{DS}}$ are the theoretically predicted energy densities of the CP and the DS phases at inverse temperature $\beta$.

We have performed this escaping dynamics on the $60$ equilibrium initial DS configurations, all obtained for a single graph instance of size $N$ with structural parameters $K=5$ and $K_{\textrm{ba}}=4$. From each of these starting points, we collect $1000$ first-passage escaping times by carrying out $1000$ independent runs of the escaping dynamics. We then perform data analysis on these $60000$ empirical first-passage escaping times.

Figure~\ref{fig:estimek5v4} reports the median first-passage time of the empirical results. There are strong finite-size effect. For relatively small systems (e.g., $N=180\,000$ and $N=360\,000$) the median escaping time reaches the minimum value at $\beta \approx 2.25$. When system size $N$ becomes considerably large (e.g, $N=1\,440\,000$ and $N=2\,880\,000$), the minimum value of the median escaping time gradually shifts to higher values of $\beta\approx 2.5$. These simulation results are consistent with the theoretical free energy gap curve (corresponding to $N=\infty$) shown in the inset of Fig.~\ref{fig:K5v4v1:fvsb}. Based on the same arguments as mentioned at the end of Sec.~\ref{subsec:MC},  we anticipate that as $N$ further increases the optimal value of $\beta$ will be approaching the theoretically predicted value $2.7726$.

It may be important to emphasize that the nonmonotonic phenomenon observed in Fig.~\ref{fig:estimek5v4} (associated with the emergence of spin glass phases) is qualitatively different from the nonmonotonic phenomenon of Fig.~\ref{fig:ESTK1055} (unrelated to spin glass phases).

\section{Local stability analysis}
\label{sec:lsa}

We now perform local stability analysis on the paramagnetic DS, partially polarized MP and ferromagnetic CP fixed points of the BP equation, to gain additional understanding on the free energy landscape of the planted regular random graph ensembles. 

\subsection{Incoherent perturbations}

Given a fixed-point solution $\{q_{j\rightarrow i}^*, q_{i\rightarrow j}^* \}$ of the BP equation (\ref{eq:BPoriginal}), let us perturb each cavity message $q_{j\rightarrow i}$ by a small quantity $\epsilon_{j\rightarrow i}$ such that $q_{j\rightarrow i} = q_{j\rightarrow i}^* (1 + \epsilon_{j\rightarrow i})$. The linearized BP iteration at the vicinity of the fixed point is then
\begin{equation}
  \epsilon_{j \rightarrow i}  \, \Leftarrow \,
  - \bigl( 1 - q_{j \rightarrow i}^* \bigr)
  \sum\limits_{k\in \partial j\backslash i} \epsilon_{k\rightarrow j} \; .
  \label{eq:231213a}
\end{equation}
Let us assume for the moment that the input perturbations $\epsilon_{k\rightarrow j}$ to vertex $i$ are mutually uncorrelated random variables (local perturbations toward disordered spin glass states~\cite{Rivoire-etal-2004}). Then the mean square of the perturbation $\epsilon_{j\rightarrow i}$ evolves according to
\begin{equation}
  \bigl\langle \epsilon_{j \rightarrow i}^2 \bigr\rangle  \, \Leftarrow \,
  \bigl( 1 - q_{j \rightarrow i}^* \bigr)^2
  \sum\limits_{k\in \partial j\backslash i}
  \bigl\langle \epsilon_{k\rightarrow j}^2 \bigr\rangle \; ,
  \label{eq:240724a}
\end{equation}
where $\langle \ldots \rangle$ is the averaged value over many independent trajectories of the BP evolution process.

We have $q_{j\rightarrow i}^* = q$ with $q$ determined by Eq.~(\ref{eq:qRSrr}) for the paramagnetic DS phase.  It is easy to see that the values of the mean squared perturbations $\bigl\langle \epsilon_{j\rightarrow i}^2 \bigr\rangle$ will amplify rather than shrink to zero if $(K-1) (1 - q)^2 > 1$. We can get a critical inverse temperature $\beta_{\textrm{pg}}$ by obtaining the unique positive root of the following equation
\begin{equation}
  1 - q \, = \, \frac{1}{\sqrt{K-1} } \; .
  \label{eq:betasg}
\end{equation}
When the inverse temperature $\beta$ exceeds $\beta_{\textrm{pg}}$, it is no longer possible for the paramagnetic DS phase to remain its ergodic property. The critical value $\beta_{\textrm{pg}}$ is an upper-bound inverse temperature beyond which many spin glass macroscopic states must form within the subspace of disordered microscopic configurations. Notice that $\beta_{\textrm{pg}}$ is independent of the inter-group degree $K_{\textrm{ba}}$. For vertex degree $K=5$ we have $\beta_{\textrm{pg}}=2.7726$ with the associated energy density $\rho = 0.6667$, while $\beta_{\textrm{pg}} = 2.9560$ with the associated energy density $\rho = 0.75$ for the ensembles of degree $K=10$.

The lowest inverse temperature to experience ergodicity breaking in the paramagnetic DS phase and the emergence of many macroscopic spin glass states is the spin glass dynamical phase transition point $\beta_d$~\cite{Mezard-Montanari-2009,Mezard-Montanari-2006}. The value of $\beta_d$ can be determined by population dynamics simulations (Appendix~\ref{app:dsgt}). According to Ref.~\cite{Zhang-Zeng-Zhou-2009}, $\beta_d$ coincides with the local instability point ($\beta_d = \beta_{\textrm{pg}}$) for regular random graph ensembles with degree $K < 16$ including the $K=10$ and $K=5$ ensembles studied here.  When the vertex degree $K \geq 16$, the DS phase will still be locally stable towards spin glass states when the spin glass dynamical phase transition occurs, namely $\beta_d < \beta_{\textrm{pg}}$~\cite{Zhang-Zeng-Zhou-2009}. For such higher-degree graph ensembles the local instability point $\beta_{\textrm{pg}}$ will then be less relevant than the dynamical transition point $\beta_d$.

\subsection{Coherent perturbations}

The linearized discrete BP iteration process (\ref{eq:231213a}) indicates the following linearized continuous time evolution process,
\begin{equation}
  \frac{ {\rm d} \epsilon_{j\rightarrow i}}{ {\rm d} t}
  \, = \,  - \epsilon_{j\rightarrow i} - (1 - q_{j\rightarrow i}^*)
  \sum\limits_{k\in \partial j\backslash i} \epsilon_{k\rightarrow j} \; .
  \label{eq:231216a}
\end{equation}
The first term on the right-hand side is the spontaneous decay, and the second term is the driving force from the adjacent vertices $k$. For regular random graph ensembles with a planted vertex cover solution, let us consider coherent perturbations of the following form
\begin{equation}
  \epsilon_{j\rightarrow i} \, = \,
  \left\{
  \begin{array}{ll}
    \epsilon_{\textrm{ab}}  \quad \quad &  ( j \in A \; ,\, i \in B ) \; , \\
    \epsilon_{\textrm{ba}}  \quad \quad &  ( j \in B \; ,\, i \in A ) \; ,\\
    \epsilon_{\textrm{bb}}  \quad \quad &  ( j \in B \; ,\, i \in B ) \; ,
  \end{array}
  \right.
\end{equation}
where $\epsilon_{\textrm{ab}}$, $\epsilon_{\textrm{ba}}$, and $\epsilon_{\textrm{bb}}$ are the three group-dependent perturbation variables. The eigenmodes of the perturbative evolution equation (\ref{eq:231216a}) are then determined by the eigenvalue problem with eigenvalues $\lambda$,
\begin{widetext}
  \begin{equation}
    \begin{bmatrix}
      -1 &  - (K - 1) (1 - q_{\textrm{ab}}) & 0   \\
      - (K_{\textrm{ba}}-1) (1 - q_{\textrm{ba}} ) & -1 & - K_{\textrm{bb}}
      (1 - q_{\textrm{ba}}) \\
      - K_{\textrm{ba}} (1 - q_{\textrm{bb}} ) & 0 & - 1 - (K_{\textrm{bb}} - 1 )
      (1 - q_{\textrm{bb}} ) 
    \end{bmatrix}
    \, \begin{bmatrix}
      \epsilon_{\textrm{ab}} \\
      \epsilon_{\textrm{ba}} \\
      \epsilon_{\textrm{bb}}
    \end{bmatrix}
    \, = \,
    \lambda \,
    \begin{bmatrix}
      \epsilon_{\textrm{ab}} \\
      \epsilon_{\textrm{ba}} \\
      \epsilon_{\textrm{bb}}
    \end{bmatrix}
    \; ,
    \label{eq240912a}
  \end{equation}
\end{widetext}
where $q_{\textrm{ab}}$, $q_{\textrm{ba}}$ and $q_{\textrm{bb}}$ satisfy Eq.~(\ref{eq:BPrr}).

One of the eigenvalues ($\lambda_1$) of the above equation is always real and negative. The corresponding eigenvector is characterized by $\epsilon_{\textrm{ab}}$, $\epsilon_{\textrm{ba}}$ and $\epsilon_{\textrm{bb}}$ all being positive (or negative). Since $\lambda_1 < 0$, this eigenmode will decay with time. We therefore focus on the other two eigenvalues $\lambda_2$ and $\lambda_3$ in the following discussions.

Consider the paramagnetic DS fixed point, $q_{\textrm{ab}} = q_{\textrm{ba}} = q_{\textrm{bb}} = q$. The eigenvalue problem (\ref{eq240912a}) then reveals the local stability of the DS phase with respect to the ferromagnetic CP phase. The three eigenvalues are
\begin{equation}
  \begin{aligned}
    \lambda_1 & \, = \, - 1 - (1 - q) (K-1) \; , \\
    \lambda_2 & \, = \,  - 1 + (1 - q) \frac{K_{\textrm{ba}} +
      \sqrt{K_{\textrm{ba}}^2 - 4 (K-1)}}{2}
    \; , \\
    \lambda_3 & \, = \, - 1 + (1 - q) \frac{K_{\textrm{ba}} - \sqrt{K_{\textrm{ba}}^2
        - 4 (K-1)}}{2}
    \; .
  \end{aligned}
\end{equation}

If $K_{\textrm{ba}}^2 > 4 (K-1)$, there is a critical inverse temperature $\beta_{\textrm{pf}}$ as determined by
\begin{equation}
  1 - q \, = \, \frac{2}{K_{\textrm{ba}} + \sqrt{ K_{\textrm{ba}}^2 - 4 (K - 1)}} \; ,
\end{equation}
at which $\lambda_2$ changes from being negative to being positive (the DS phase becomes locally unstable). Because of the fact that 
\begin{equation}
  \frac{2}{K_{\textrm{ba}} + \sqrt{K_{\textrm{ba}}^2 - 4 (K-1)}} \, < \,
  \frac{1}{\sqrt{K-1}} \; ,
\end{equation}
we see that $\beta_{\textrm{pf}} < \beta_{\textrm{pg}}$. This means that the instability toward the ferromagnetic CP phase occurs in the DS phase before the local instability toward the spin glass phases. The ensemble of $K=10$, $K_{\textrm{ba}} = 7$ discussed in Section~\ref{sec:10_7_3} is one example of this catalog.

On the other hand, if $K_{\textrm{ba}}^2 < 4 (K - 1)$, the eigenvalues $\lambda_2$ and $\lambda_3$ form a complex conjugate pair. Then the instability point $\beta_{\textrm{pf}}$ of the DS phase with respect to the ferromagnetic CP phase is determined by
\begin{equation}
  1 - q \, = \, \frac{2}{K_{\textrm{ba}}} \; .
\end{equation}
Because $K_{\textrm{ba}} <  2 \sqrt{K-1}$, we have $\beta_{\textrm{pf}} > \beta_{\textrm{pg}}$, and therefore the paramagnetic DS phase will still be locally stable with respect to the ferromagnetic CP phase when it becomes locally unstable with respect to the spin glass phases. The ensemble of $K_{\textrm{ba}} = 5$ and $K = 10$ discussed in Section~\ref{sec:10_5_5} is one example of this catalog.

If $K_{\textrm{ba}}^2 = 4 (K-1)$ exactly, the two eigenvalues $\lambda_2$ and $\lambda_3$ are degenerate and they become positive when $(1-q) \sqrt{K-1} > 1$. This means that the critical inverse temperature $\beta_{\textrm{pf}}$ is also determined by Eq.~(\ref{eq:betasg}), and hence it is identical to $\beta_{\textrm{pg}}$. This property of $\beta_{\textrm{pf}}= \beta_{\textrm{pg}}$ means that the paramagnetic DS phase will become locally unstable both towards the ferromagnetic CP phase and toward the spin glass phases simultaneously. The ensembles of $K=10$, $K_{\textrm{ba}} = 6$ discussed in Section~\ref{sec:10_6_4} and the one of $K=5$, $K_{\textrm{ba}} = 4$ of Section~\ref{sec:5_4_1} are two examples of this catalog. It is easy to verify that the condition $K_{\textrm{ba}}^2 = 4 (K-1)$ is equivalent to the condition (\ref{eq:Kscaling}), and therefore all the special regular random graph ensembles discussed in Sec.~\ref{sec:special} have $\beta_{\textrm{pf}} = \beta_{\textrm{pg}}$.

\begin{figure}[b]
  \centering
  \hspace*{-0.15cm}
  \subfigure[]{
    \includegraphics[angle=270, width=0.49\linewidth]{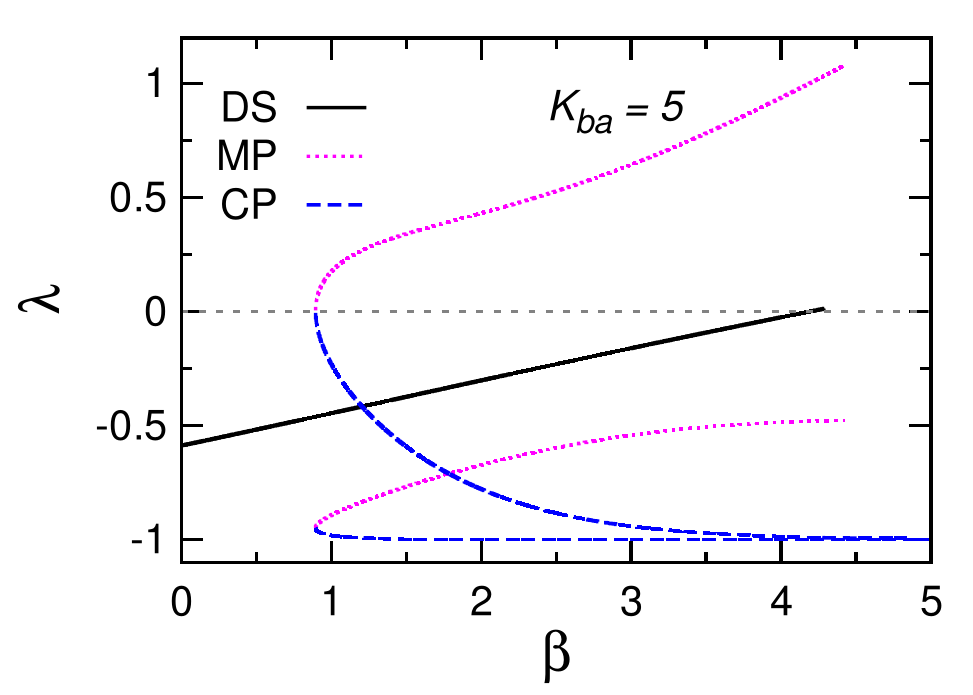}
    \label{fig:BPstabK1055}
  } \hspace*{-0.4cm}
  \subfigure[]{
    \includegraphics[angle=270, width=0.49\linewidth]{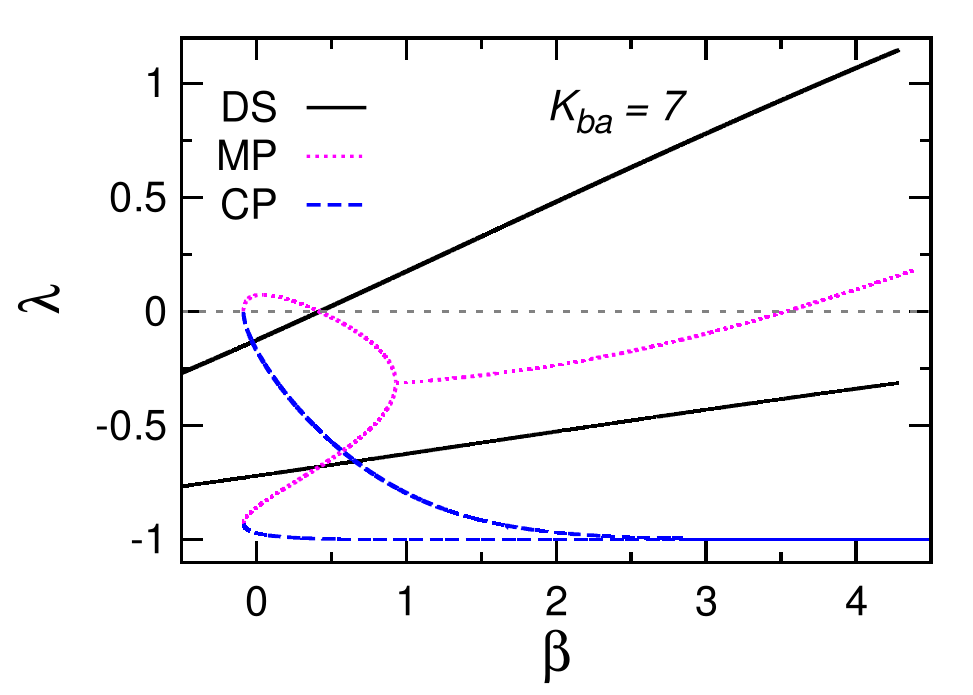}
    \label{fig:BPstabK1073}
  }
  \\
  \subfigure[]{
    \includegraphics[angle=270, width=0.49\linewidth]{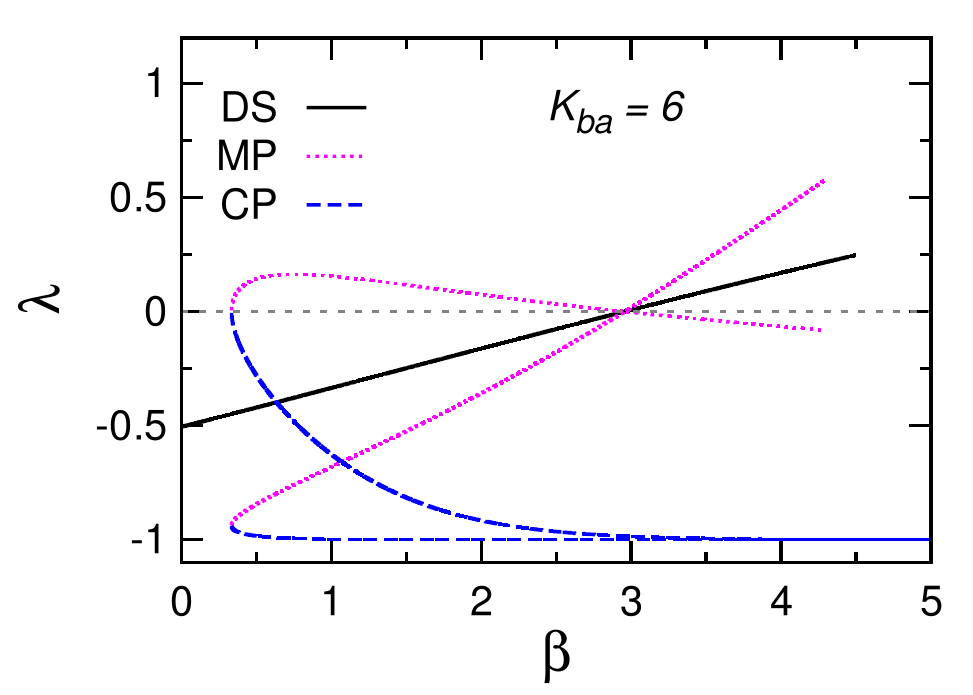}
      \label{fig:BPstabK1064}
  }
  \caption{
    Local stability analysis on the three planted regular random graph ensembles. The vertex degree is $K=10$ and the inter-group degree $K_{\textrm{ba}}$ is $5$ (a), $7$ (b), and $6$ (c). We plot the two non-trivial eigenvalues $\lambda_2$ and $\lambda_3$ of the local stability eigenvalue problem (\ref{eq240912a}) for the DS, MP, and CP fixed points. If $\lambda_2$ and $\lambda_3$ form a complex conjugate pair, only the real part is shown. 
  }
  \label{fig:BPstabK10}
\end{figure}

We perform the same local stability analysis on the MP and CP fixed points of the BP equation and obtain the eigenvalues $\lambda_2$ and $\lambda_3$ of the corresponding eigenvalue problems (\ref{eq240912a}). Figure~\ref{fig:BPstabK10} summarizes the numerical results for the three representative graph ensembles of degree $K=10$. These results confirm that the ferromagnetic CP phase is always locally stable. Concerning the MP fixed point, we find that it is always locally unstable if $K_{\textrm{ba}}=5$ (Fig.~\ref{fig:BPstabK1055}) or $K_{\textrm{ba}}=6$ (Fig.~\ref{fig:BPstabK1064}), and it changes from being locally unstable to being locally stable at the critical inverse temperature $\beta_{\textrm{pf}}=0.4213$ if $K_{\textrm{ba}}=7$ (Fig.~\ref{fig:BPstabK1073}). These are consistent with the free energy results of the preceding sections.

\section{Belief-propagation evolution dynamics}
\label{sec:bpcted}

When the cavity messages $q_{j\rightarrow i}$ are far away from a fixed point of the BP equation (\ref{eq:BPoriginal}), the linearized evolution process (\ref{eq:231216a}) is no longer appropriate, but we can still write a nonlinear continuous-time evolution equation as
\begin{equation}
  \frac{ \textrm{d} q_{j\rightarrow i}(t)}{\textrm{d} t} \, = \,
  - q_{j\rightarrow i}(t) +
  \frac{e^{-\beta}}{e^{-\beta} + \prod\limits_{k \in \partial i\backslash j}
    q_{k\rightarrow i}(t)} \; .
  \label{eq240913a}
\end{equation}

Starting from a random initial condition, the evolution trajectory of the high-dimensional vector $\{q_{j\rightarrow i}(t)\}$ with time $t$ will bring us valuable information concerning the free energy landscape of the planted vertex cover problem. If this evolutionary trajectory has a high probability of converging to the ferromagnetic CP phase at inverse temperatures $\beta > \beta_{\textrm{pf}}$, it  can serve as a simple algorithm for solving the planted vertex cover problem. This is an interesting and significant topic for further studies (see Ref.~\cite{Angelini-RicciTersenghi-2022} for related discussions).

Here we use Eq.~(\ref{eq240913a}) to gain further insights into the free energy landscapes of type Fig.~\ref{fig:LMr}. Especially we wish to understand: Why the paramagnetic DS phase for a special graph ensemble with $K$ and $K_{\textrm{ba}}$ obeying Eq.~(\ref{eq:Kscaling}) will be locally unstable at $\beta > \beta_{\textrm{pf}}$ with respect to the ferromagnetic CP phase, when it is protected by an intermediate MP phase with positive magnetization $m$ and positive free energy density gap $\Delta f$?

To investigate this issue, we restrict the evolution (\ref{eq240913a}) to the coherent group-dependent dynamics involving only three cavity messages $q_{\textrm{ab}}$(t), $q_{\textrm{ba}}(t)$ and $q_{\textrm{bb}}(t)$. The corresponding three-dimensional evolution equation is
\begin{equation}
  \begin{aligned}
    \frac{ \textrm{d} q_{\textrm{ab}}}{ \textrm{d} t} \, & = \,
    - q_{\textrm{ab}} + \frac{e^{-\beta}}{e^{-\beta} + (q_{\textrm{ba}})^{K-1}} \; ,
    \\
    \frac{ \textrm{d} q_{\textrm{ba}}}{ \textrm{d} t} \, & = \,
    - q_{\textrm{ba}} + \frac{e^{-\beta}}{e^{-\beta}
      + (q_{\textrm{ab}})^{K_{\textrm{ba}} - 1} (q_{\textrm{bb}})^{K- K_{\textrm{ba}}}} \; ,
    \\
    \frac{ \textrm{d} q_{\textrm{bb}}}{ \textrm{d} t} \, & = \,
    - q_{\textrm{bb}} + \frac{e^{-\beta}}{e^{-\beta} + (q_{\textrm{ab}})^{K_{\textrm{ba}}}
      (q_{\textrm{bb}})^{K- K_{\textrm{ba}}-1}} \; . 
  \end{aligned}
  \label{eq:240902b}
\end{equation}
Starting from a random initial condition which is very close to the paramagnetic DS fixed point, the distance of the evolution trajectory at time $t$ from this DS fixed point is computed through
$$
\Delta_{D S} \, = \, 
\sqrt{ (q_{\textrm{ab}}(t) - q)^2 + (q_{\textrm{ba}}(t) - q)^2
  + (q_{\textrm{bb}}(t) - q)^2 } \; ,
$$
and the corresponding distances $\Delta_{\textrm{MP}}$ and $\Delta_{\textrm{CP}}$ to the MP and CP fixed points are defined in the same way.

\begin{figure}
  \centering
  \includegraphics[angle=270,origin=c,width=0.42\linewidth]{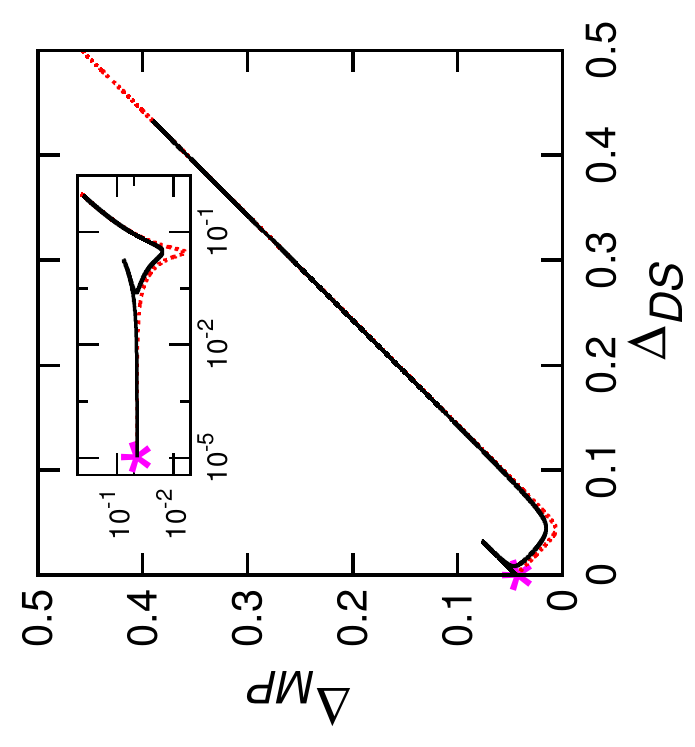}
  \hspace{0.2cm}
  \raisebox{0.45cm}{
    \includegraphics[width=0.45\linewidth]{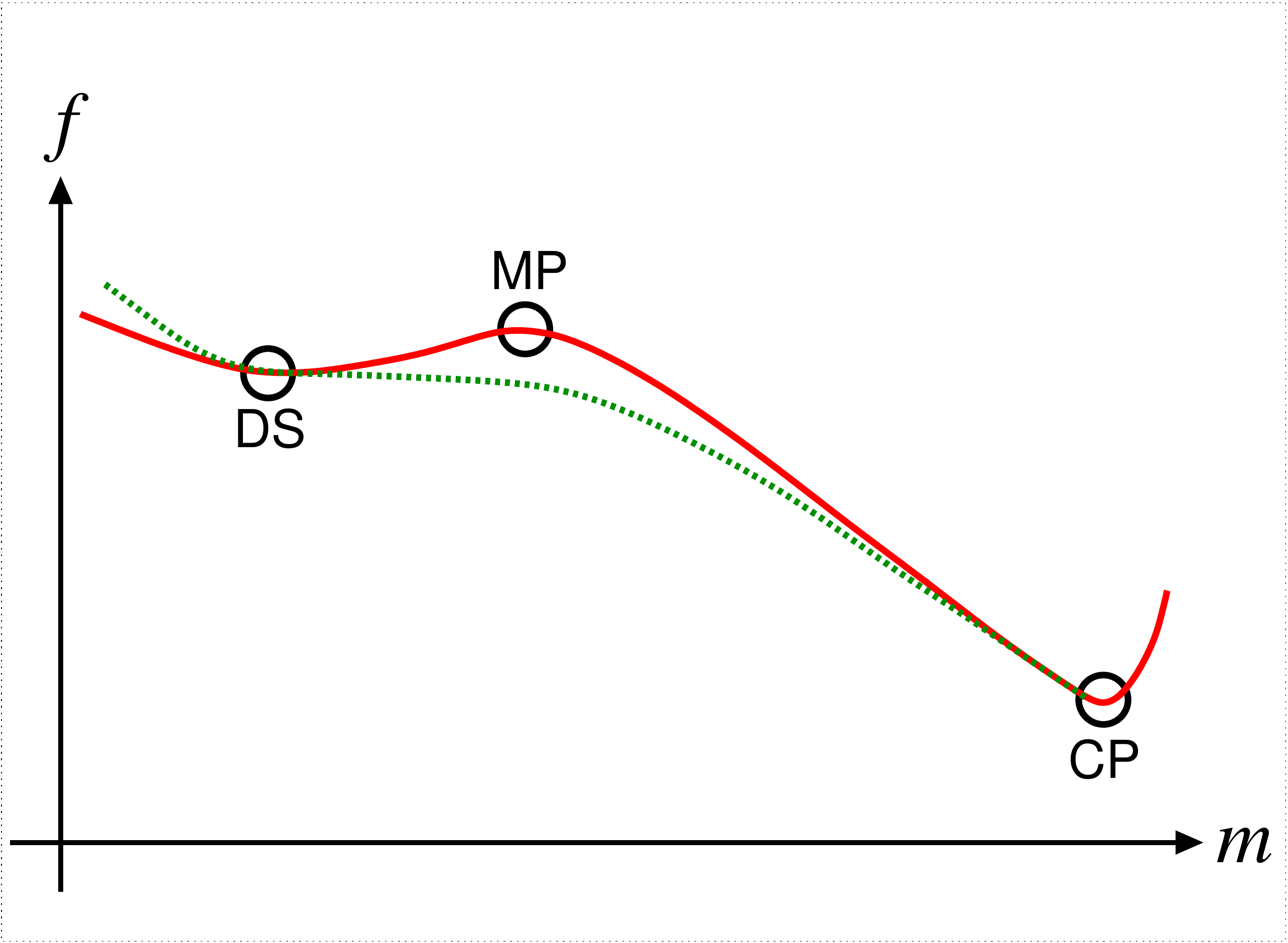}
  }
  \caption{
    Belief-propagation evolution away from the paramagnetic DS phase in the random graph ensemble with degree parameters $K=5$, $K_{\textrm{ba}}=4$. (left) Distances $\Delta_{\textrm{DS}}$ and $\Delta_{\textrm{MP}}$ to the DS and MP fixed points at $\beta = 4.0$, starting from two slightly different initial conditions located in the vicinity of the star symbol. Inset is the log-log plot. (right) Schematic free energy landscape at $\beta = 4.0$, with magnetization $m$ as the order parameter. The solid and dotted curves indicate MP $\rightarrow$ CP and DS $\rightarrow$ CP evolution trajectories which are also indicated in Fig.~\ref{fig:LMr}.
  }
  \label{fig:BPevolution}
\end{figure}

Figure~\ref{fig:BPevolution} (left) shows two representative evolution trajectories, obtained at $\beta = 4.0$ for the special graph ensemble of $K=5$, $K_{\textrm{ba}}=4$ discussed in Sec.~\ref{sec:5_4_1}. In one type of typical trajectories, we find that $\Delta_{\textrm{DS}}$ increases monotonically with time $t$  while $\Delta_{\textrm{MP}}$ first decreases with time and reaches a small minimum value and then increases monotonically with time, implying that the evolution process is steadily escaping away from the DS phase and reaching the ferromagnetic CP phase after passing through the neighborhood of the MP fixed point. In the other type of trajectories, we notice that the trajectory first deviate away from both the DS and the MP fixed points and then is dragged back to the vicinity of the DS fixed point, and finally it escapes the DS region and be absorbed to the ferromagnetic CP phase after passing through the neighborhood of the MP fixed point.

These simulation results confirm that the paramagnetic DS fixed point at $\beta > \beta_{\textrm{pf}}$ is indeed only a saddle point of the ferromagnetic free energy landscape. We may schematically illustrate the basin of attraction of the ferromagnetic CP phase at $\beta > \beta_{\textrm{pf}}$ through Fig.~\ref{fig:BPevolution} (right). Both the DS and the MP fixed points mark saddle points of the free energy landscape. Although the MP fixed point has higher free energy density than the DS fixed point, it fails to protect the paramagnetic DS phase, as there are paths linking the DS fixed point directly to the CP fixed point and bypassing the MP one.

Similar free energy landscape property of the form Fig.~\ref{fig:BPevolution} (right) may also be observed for the graph ensemble of $K=10$, $K_{\textrm{ba}}=7$ for which the MP fixed point and the DS fixed point both become unstable at large values of $\beta > 3.5$, see Fig.~\ref{fig:BPstabK1073}.  The anti-ferromagnetic MP phase of this system will then disappear at such high inverse temperatures (low temperatures).

\section{Conclusion and outlook}
\label{sec:co}

We planted a minimum vertex cover in a regular random graph and investigated the free energy landscape of this inference and optimization system by the cavity method, and we carried out numerical experiments on the escaping dynamics from the paramagnetic DS phase towards the planted ferromagnetic CP phase. The two degree parameters of the planted graph ensemble are the vertex degree $K$ and the inter-group degree $K_{\textrm{ba}}$. This planted vertex cover problem has a discontinuous equilibrium phase transition between the paramagnetic DS phase and the ferromagnetic CP phase. The critical inverse temperatures at which the paramagnetic DS phase becomes unstable towards the ferromagnetic CP phase ($\beta_{\textrm{pf}}$) and becomes unstable towards disordered spin glass phases ($\beta_{\textrm{pg}}$) have been analytically determined. We found that there are three distinct types of free energy landscapes which have the property of $\beta_{\textrm{pf}} < \beta_{\textrm{pg}}$ valid for $K_{\textrm{ba}}^2 > 4 (K-1)$, and $\beta_{\textrm{pf}} > \beta_{\textrm{pg}}$ valid for $K_{\textrm{ba}}^2 < 4 (K-1)$, and $\beta_{\textrm{pf}} = \beta_{\textrm{pg}}$ only for $K_{\textrm{ba}}^2 = 4 (K-1)$, respectively.

For graph ensembles with $K_{\textrm{ba}}^2 > 4 (K-1)$ and hence $\beta_{\textrm{pf}} < \beta_{\textrm{pg}}$, we found that a locally stable anti-ferromagnetic MP phase starts to emerge in the free energy landscape at the critical inverse temperature $\beta_{\textrm{pf}}$.

If $K_{\textrm{ba}}^2 < 4 (K-1)$ and hence $\beta_{\textrm{pf}} > \beta_{\textrm{pg}}$, we discovered that the rescaled free energy barrier $\beta \Delta f$ attains a positive minimum value at certain inverse temperature $\beta_{opt} < \beta_{\textrm{pf}}$ (see Fig.~\ref{fig:K10v5v5:dbf}). This nonmonotonic thermodynamic property implies nonmonotonic mean first-passage time of escaping the paramagnetic DS phase, and this prediction was consistent with computer simulation results on single graph instances with degree parameters $K=10$, $K_{\textrm{ba}}=5$ (see Fig.~\ref{fig:ESTK1055}).

For planted graph ensembles with $K_{\textrm{ba}}^2 \geq 4 (K-1)$ and hence $\beta_{\textrm{pf}} \leq \beta_{\textrm{pg}}$, as the stochastic local search dynamics running at $\beta > \beta_{\textrm{bf}}$ may be trapped by the anti-ferromagnetic MP phase or by the disordered spin glass phases, our mean field theory also implies that the mean first-passage time of escaping the paramagnetic DS phase will be a nonmonotonic function of inverse temperature. This prediction was again supported by computer simulation results obtained on single graph instances with degree parameters $K=5$, $K_{\textrm{ba}}=4$ (see Fig.~\ref{fig:estimek5v4}).

These theoretical results therefore indicate that there are at least two distinct thermodynamic reasons behind the nonmonotonic temperature dependence in the supercooled region. 

To extend the present work, it may be interesting to relax the hard constraint (\ref{eq240919a}) and replace it with a soft constraint with a finite energy penalty~\cite{Dote-Hukushima-2024,Dote-Hukushima-2024b}. This modification may have nontrivial effect on the free energy landscape property. It remains to be checked whether the discontinuous nature of the equilibrium ferromagnetic phase transition will be affected. 

Another immediate extension of this work is to check whether the nonmonotonic properties (Fig.~\ref{fig:K10v5v5diff}) also hold for general planted random-graph glass models in which an empty vertex can have up to $z \geq 1$ empty nearest neighbors~\cite{Biroli-Mezard-2002}. We are working on this issue and will present the quantitative results in a follow-up publication.

As the planted vertex cover problem contains only two-body interactions and is restricted to binary vertex states, it may be relatively easy to implement this model in quantum computing circuits~\cite{Bellitti-etal-2021,Zeng-etal-2024,Bapst-etal-2013,Bernaschi-etal-2021}. As this model system has a discontinuous ferromagnetic phase transition, it may be useful as a challenging benchmark problem for quantum optimization algorithms which usually work at nearly zero temperature~\cite{Kadowaki-Nishimori-1998}. 

Besides serving as an interesting inference problem, the planted vertex cover problem can also be regarded as a model of glass-forming liquids with a planted crystalline state~\cite{Biroli-Mezard-2002,Rivoire-etal-2004}. Although it has been known for a long time that the vertex cover problem defined on some two-dimensional and three-dimensional lattices has a continuous crystallization phase transition~\cite{Baxter-1980,Gaunt-1967}, it is still very interesting to explore the possibility of a discontinuous crystallization phase transition even in two-dimensional and three-dimensional planted graph systems with two or more groups of vertices, and to check whether the mean first-passing time to reach the crystalline phase is also a nonmonotonic function of inverse temperature in such systems. The vertex cover problem is a limiting case of the $K$-core attack problem, which is itself closely linked to the Fredrickson-Andersen kinetically constrained spin model~\cite{Zhou-2024}. Along this direction, a future pursuit will be the extension of the planted vertex cover problem to the planted $K$-core attack problem. 

The nonmonotonic temperature dependence of the median waiting time $\tau$ (Fig.~\ref{fig:ESTK1055} and Fig.~\ref{fig:estimek5v4}) observed in our computer simulations may be the joint consequence of thermodynamic and kinetic effects. We have only considered the thermodynamic effect associated with $\beta \Delta f$. An empty vertex will temporarily block all its nearest neighbors from shifting occupation state. This kinetic effect is stronger at higher inverse temperature values $\beta$ as more vertices become empty. Quantifying the kinetic effect to $\tau$ might be a very challenging task. On the thermodynamic side, we know from Eq.~(\ref{eq:barrierslope}) that, when $\beta < \beta_{opt}$ the escaping dynamics is facilitated by energy and when $\beta > \beta_{opt}$ it is hindered by energy, and at $\beta= \beta_{opt}$ the escaping dynamics is purely entropic in nature. This insight might also be relevant for real-world (three-dimensional) supercooled liquids, whose nonmonotonic crystallization behavior arises from competing ordered and disordered structures~\cite{Schmelzer-etal-2016}. We hope our present work will stimulate more future work on supercooled liquids in random graph models and in finite-dimensional lattices.

\begin{acknowledgments}
  We thank Prof.~Federico Ricci-Tersenghi for sending us a helpful comment on an earlier version of this manuscript. The following funding supports are acknowledged: National Natural Science Foundation of China Grants No.~12247104 and No.~12047503. Numerical simulations were carried out at the HPC cluster of ITP-CAS and also at the BSCC-A3 platform of the National Supercomputer Center in Beijing.
\end{acknowledgments}

\begin{appendix}

\section{Monte Carlo simulation details}
\label{app:mcmc}

There are two types of elementary operations to update a microscopic occupation configuration $\vec{\bm{c}}$. The first type is flipping a single vertex $i$ from its old state $c_i$ to the complementary state $c_i^\prime = 1 - c_i$. A vertex $i$ is first chosen uniformly at random from the whole set of $N$ vertices. To obey detailed balance, if the old state $c_i = 0$, the flip is accepted with probability $e^{- \beta}$ and is rejected with the remaining probability $1- e^{-\beta}$. If the old state $c_i = 1$, the flip is accepted if all the adjacent vertices $j$ of this vertex are occupied ($c_j = 1$) at the moment, otherwise the flip  is rejected and vertex $i$ keeps the old state $c_i$.

The second type is swapping the states of two vertices $i$ and $j$. It goes as follows: (1) Randomly pick up an unoccupied vertex $i$ and flip it ($c_i: 0 \rightarrow 1$); (2) then randomly pick up a vertex $j$ from the whole set of occupied and flippable vertices and change its state to be empty ($c_j: 1\rightarrow 0$). This swapping process means random (and possibly long-distance) hopping of a particle from one site to another unoccupied site, without changing the number of occupied vertices (the energy). This swapping dynamics has shown to be very efficient for sampling low-temperature equilibrium configurations for structural glasses~\cite{Berthier-etal-2018}.

The swapping trials are not essential for our numerical experiments, they are introduced only for the purpose of reaching equilibrium within the paramagnetic DS phase more quickly. We set the probability of performing the swapping process to be $0.9$ and the probability of performing the single-vertex flipping process to be $0.1$. Vertex cover configurations $\vec{\bm{c}}$ of the paramagnetic DS phase do not contain information about the planted configuration. The mean fraction $\rho_{\textrm{a}}$ of occupied vertices in group $A$ is equal to the fraction $\rho_{\textrm{b}}$ of occupied vertices in group $B$. To ensure that the evolution trajectory stays in the DS phase, if a single-vertex flipping trial or a swapping trial will enlarge the difference between $\rho_{\textrm{a}}$ and $\rho_{\text{b}}$, we reject such trials. By this way we find that the $\rho_{\textrm{a}}$ and $\rho_{\textrm{b}}$ values are fluctuating slightly around the same value during the equilibration process.

We sample $60$ independent equilibrium microscopic configurations from the paramagnetic DS phase through the above-mentioned method. Starting from each of these initial configurations, we then simulate $1000$ independent evolution trajectories using only single-vertex flipping trials and no longer restricting the trajectories to be within the paramagnetic DS phase. One time step of the Monte Carlo simulation then corresponds to $N$ single-vertex flipping trials. The first-passage time of each evolution trajectory to reach the ferromagnetic CP phase is stored. We take the median value of these $60000$ first-passage escaping time as the characteristic waiting time of escaping the paramagnetic DS phase.

If the energy density $\rho$ is fixed instead of the inverse temperature $\beta$ (microcanonical simulations), we apply both single-vertex flipping and vertex-pair swapping trials in the evolution dynamics. An energy upper-bound is set as  $N \rho$. If a proposed change of the microscopic configuration will make the energy go beyond this upper-bound, such a  proposal is rejected. The inverse temperature $\beta$ is then estimated following the same protocol as in Refs.~\cite{Zhou-2019,Zhou-Liao-2021}.

\section{Spin glass dynamical phase transition}
\label{app:dsgt}
  
The spin glass dynamical phase transition point $\beta_d$ is defined as the critical inverse temperature at which the ergodic property of the paramagnetic DS phase starts to be severely broken and a huge number of clusters of microscopic configurations suddenly emerge in the DS configuration space~\cite{Mezard-Montanari-2006}. These clusters are referred to as macroscopic states. It has been shown that $\beta_d$ coincides with the spin glass local instability point as determined by Eq.~(\ref{eq:betasg}) for regular random graph ensembles with degree $K < 16$~\cite{Zhang-Zeng-Zhou-2009}. To verify that this conclusion also holds for the planted regular graph ensembles, we adapt the method of Ref.~\cite{Zhang-Zeng-Zhou-2009} to our planted systems with two degree parameters $K$ and $K_{\textrm{ba}}$. Here we describe the technical details. The computation assumes that the inverse temperatures at the level of macroscopic states and at the level of microscopic configurations are equal to each other.

When there are a huge number of spin glass macroscopic states, the cavity message $q_{i\rightarrow j}$ from vertex $i$ to vertex $j$ will be different for different macroscopic states. Let us denote by $Q_{i\rightarrow j}^{0}[ q_{i\rightarrow j} ]$ the conditional probability of $q_{i\rightarrow j}$ in a macroscopic state $\alpha$, given that the sampled state of vertex $i$ from this macroscopic state is $c_i = 0$. Similarly the conditional probability $Q_{i\rightarrow j}^{1}[ q_{i\rightarrow j} ]$ conditional on $c_i = 1$ can be defined on edge the $(i, j)$. When all the microscopic configurations of all these macroscopic states are combined together, the paramagnetic DS phase should be recovered, therefore the mean value of $q_{i\rightarrow j}$ among all these macroscopic states  should be equal to $q$, the root of Eq.~(\ref{eq:qRSrr}).

If we observe $c_i = 0$, all the adjacent vertices $k$ of vertex $i$ must all be in the occupied state ($c_k = 1$). This means the following self-consistent equation,
\begin{equation}
  \begin{aligned}
    Q_{i\rightarrow j}^0[ q_{i\rightarrow j} ] \,
    = \, & \prod\limits_{k\in \partial i\backslash j} \int \textrm{d} q_{k\rightarrow i} Q_{k\rightarrow i}^1[ q_{k\rightarrow i} ]\, \times \\
    &  \quad
    \delta\Bigl( q_{i\rightarrow j}  - \frac{ e^{-\beta} }{ e^{-\beta} +
      \prod\limits_{k\in \partial i\backslash j} q_{k\rightarrow i} } \Bigr) \; .
  \end{aligned}
  \label{eq:Q0profile}
\end{equation}
On the other hand, if $c_i = 1$ all the adjacent vertices $k$ of vertex $i$ are free to take $c_k = 1$ (with probability $q$) or $c_k = 0$ (with probability $1-q$) without violating the constraints associated with the edges $(i, k)$.  Then we obtain
\begin{equation}
  \begin{aligned}
    & Q_{i\rightarrow j}^1 [ q_{i\rightarrow j} ] =  \\
    &  \quad 
    \prod\limits_{k\in \partial i\backslash j} \Bigl[ \sum\limits_{c_k}
      q^{c_k} (1-q)^{1-c_k} \int {\rm d} q_{k\rightarrow i}
      Q_{k\rightarrow i}^{c_k}[ q_{k\rightarrow i} ] \Bigr]
    \\
    &  \quad \quad \quad 
    \times \ \delta\Bigl( q_{i\rightarrow j} -
    \frac{ e^{-\beta} }{ e^{-\beta} + \prod\limits_{k\in \partial i\backslash j}
      q_{k\rightarrow i} } \Bigr) \; .
  \end{aligned}
  \label{eq:Q1profile}
\end{equation}

For the graph ensembles studied here with random connection patterns and with every vertex having the same number of adjacent vertices, it turns out that the conditional probability functions $Q_{i\rightarrow j}^{c_i}[ q_{i\rightarrow j} ]$ are independent of the specific edges $(i, j)$ but are only dependent on the group indices $A$ and $B$,
\begin{equation}
Q_{i\rightarrow j}^{c_i}[ q_{i\rightarrow j} ] = \left\{
\begin{array}{ll}
  Q_{\textrm{ab}}^{c_i}[ q_{i\rightarrow j} ]  \hspace{1.0cm} & \quad \quad (i\in A,\;  j \in B)  \; , \\
  & \\
  Q_{\textrm{ba}}^{c_i}[ q_{i\rightarrow j} ]  \hspace{1.0cm} & \quad \quad (i\in B, \; j \in A) \; , \\
  & \\
Q_{\textrm{bb}}^{c_i}[ q_{i\rightarrow j} ]  \hspace{1.0cm} & \quad \quad (i\in B,\; j \in B) \; .
\end{array}
\right.
\end{equation}
Therefore, we only need to consider the evolution of six probability functions, $Q_{\textrm{ab}}^0$, $Q_{\textrm{ab}}^1$, $Q_{\textrm{ba}}^0$, $Q_{\textrm{ba}}^1$, $Q_{\textrm{bb}}^0$, $Q_{\textrm{bb}}^1$.  We represent these six probability functions by six real-valued arrays in our numerical iterations, each of which contains $\mathcal{P}$ samples of the cavity probability $q_{i \rightarrow j}$.

We start from the most extreme situation of every macroscopic state $\alpha$ containing only a single microscopic configuration, and then follow the merging and coarsening process of these macroscopic states by population dynamics. Initially, all the $\mathcal{P}$ cavity probability elements $q_{i\rightarrow j}$ in each of the three arrays $Q_{\textrm{ab}}^1$, $Q_{\textrm{ba}}^1$ and $Q_{\textrm{bb}}^1$ are set to be $q_{i \rightarrow j} = 1$, and all the $\mathcal{P}$ cavity probabilities in each of the three population arrays $Q_{\textrm{ab}}^0$, $Q_{\textrm{ba}}^0$ and $Q_{\textrm{bb}}^0$ are set to be $q_{i \rightarrow j}= 0$.

In a planted regular random graph the fractions of cavity probability messages $q_{i\rightarrow j}$ from a vertex $i$ of one group to a vertex $j$ of another group are, respectively,
$$
\phi_{\textrm{ab}}   =  \frac{K_{\textrm{ba}}}{K + K_{\textrm{ba}}}\; , \, \, 
\phi_{\textrm{ba}}   =  \frac{K_{\textrm{ba}}}{K + K_{\textrm{ba}}}\; , \, \, 
\phi_{\textrm{bb}}   =  \frac{K - K_{\textrm{ba}}}{K + K_{\textrm{ba}}} \; ,
$$
such that $\phi_{\textrm{ab}} + \phi_{\textrm{ba}} + \phi_{\textrm{bb}} = 1$ (notice that $\phi_{\textrm{ab}} = \phi_{\textrm{ba}}$). At each elementary evolution step, with probability $\phi_{\textrm{ab}}$ we update the pair of populations $Q_{\textrm{ab}}^0$ and $Q_{\textrm{ab}}^1$, with probability $\phi_{\textrm{ba}}$ we update the pair of populations $Q_{\textrm{ba}}^0$ and $Q_{\textrm{bb}}^1$, and with the remaining probability $\phi_{\textrm{bb}}$ we update the pair of populations $Q_{\textrm{bb}}^0$ and $Q_{\textrm{bb}}^1$. One unit of time of this population dynamics corresponds to $\mathcal{P}$ repeats of such elementary update processes.

Now we describe the details of population update, taking the task of updating the two arrays $Q_{\textrm{ba}}^0$ and $Q_{\textrm{ba}}^1$ as an example. If an edge $(i, j)$ is between a vertex $i$ of group $B$ and a vertex $j$ of group $A$, then vertex $i$ has $K_{\textrm{bb}}$ ($=K - K_{\textrm{ba}}$) adjacent vertices $k$ in group $B$ and  $K_{\textrm{ba}} - 1$ adjacent vertices $l$ in group $A$ besides vertex $j$. To update the population array $Q_{\textrm{ba}}^0$ following Eq.~(\ref{eq:Q0profile}), we sample $K_{\textrm{bb}}$ cavity probabilities (say $q_{k\rightarrow i}$ with $k=1, \ldots, K_{\textrm{bb}}$) independently and with replacement from the array $Q_{\textrm{bb}}^1$ and then sample $K_{\textrm{ba}} - 1$ cavity probabilities (say $q_{l\rightarrow i}$ with $l=K_{\textrm{bb}} + 1, \ldots, K - 1$) independently and with replacement from the array $Q_{\textrm{ab}}^1$, and then generate a new cavity probability $q_{i\rightarrow j}$ as
\begin{equation}
  q_{i\rightarrow j} \, = \,
  \frac{e^{-\beta} }{ e^{-\beta} + \prod\limits_{k}  q_{k\rightarrow i}
    \prod\limits_{l} q_{l\rightarrow i} } \; .
  \label{eq:newqij}
\end{equation}
We replace a randomly chosen element of the population array $Q_{\textrm{ba}}^0$ with this newly generated value $q_{i\rightarrow j}$.

\begin{figure}
  \centering
  \hspace*{-0.15cm}
  \subfigure[]{
    \includegraphics[angle=270,width=0.49\linewidth]{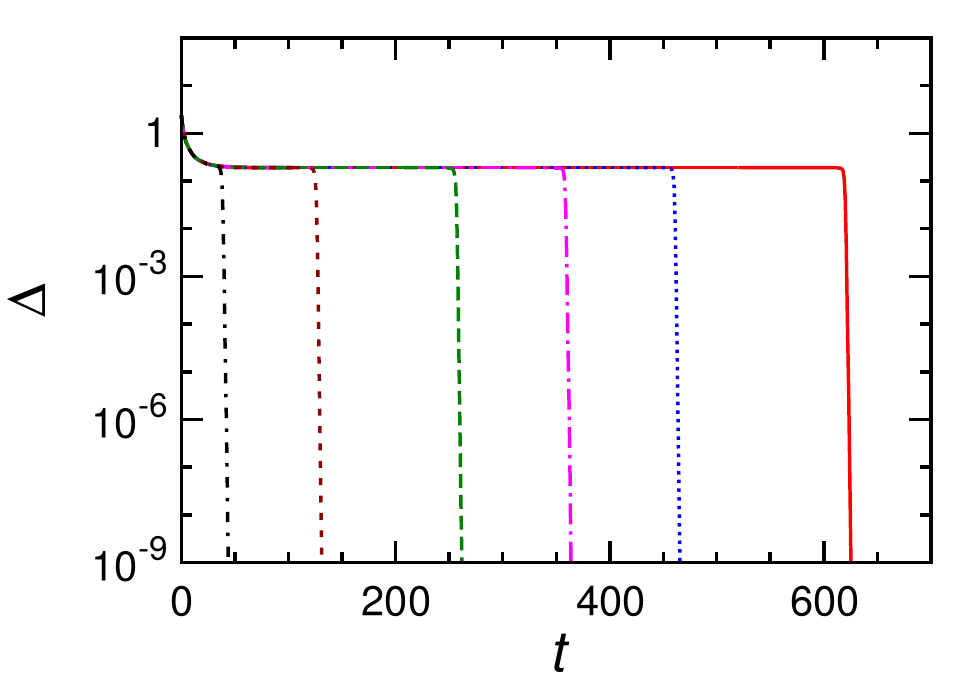}
    \label{fig:K541B2p9:delta}
  } \hspace*{-0.4cm}
  \subfigure[]{
    \includegraphics[angle=270,width=0.49\linewidth]{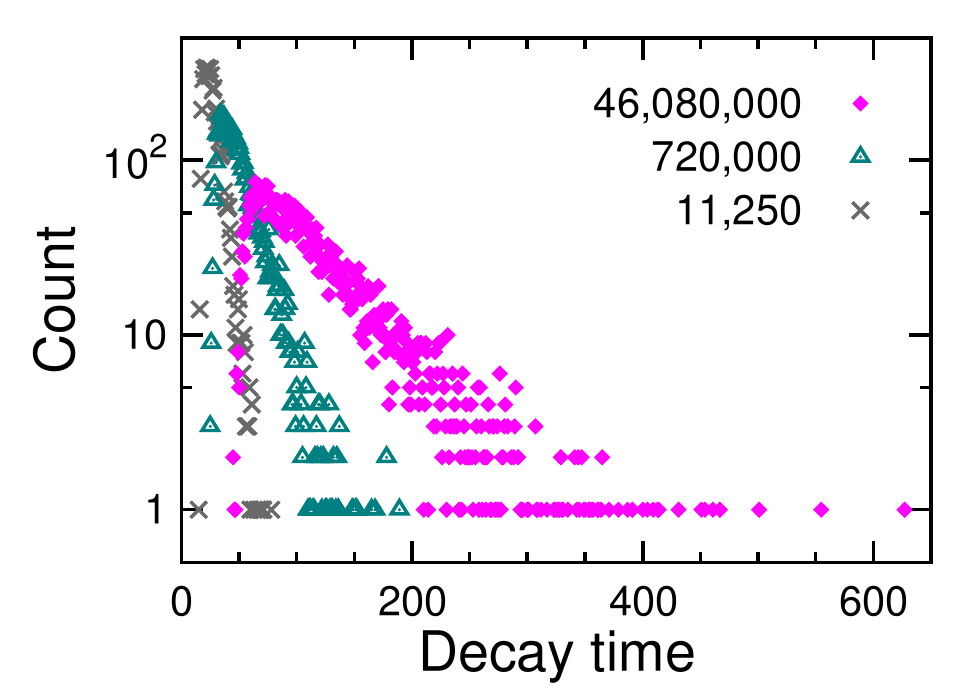}
    \label{fig:K541B2p9:tau}
  }
  \\
  \subfigure[]{
    \includegraphics[angle=270,width=0.49\linewidth]{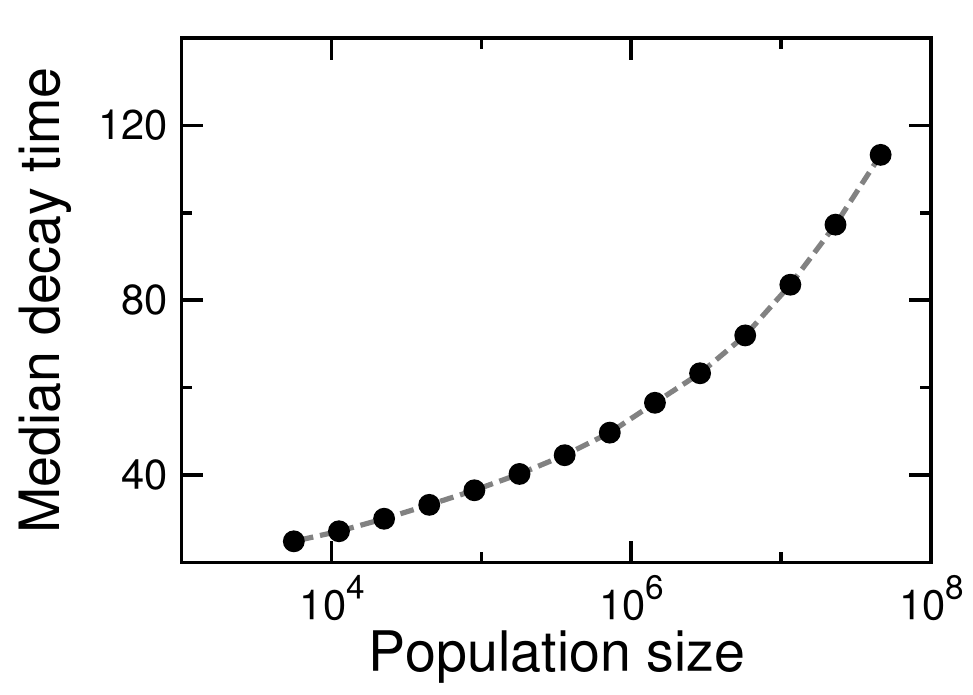}
    \label{fig:K541B2p9:scaling}
  }
  \caption{
    Population dynamics simulation results to detect spin glass macroscopic states in the planted regular random graph ensemble with degree parameters $K=5$, $K_{\textrm{ba}}=4$. (a) Example evolution trajectories for population size $\mathcal{P} = 46\,080\,000$ at $\beta = 2.9$. (b) Histograms of decay time $t_{decay}$ of sampled trajectories. Each histogram is obtained by $5040$ independent runs of the evolution dynamics at $\beta = 2.9$ with population size $\mathcal{P}=11\,250$ (cross), $720\,000$ (triangle), and $46\,080\,000$ (diamond). (c) Medium value of decay times $t_{decay}$ as a function of population size $\mathcal{P}$.}
  \label{fig:K541B2p9}
\end{figure}

On the other hand, to update the population array $Q_{\textrm{ba}}^1$ following Eq.~(\ref{eq:Q1profile}), we see that each of the $K_{\textrm{bb}}$ cavity probabilities $q_{k\rightarrow i}$ should be sampled from the population array $Q_{\textrm{bb}}^1$ only with probability $q$ and it should be sampled from the population array $Q_{\textrm{bb}}^0$ with the remaining probability $1-q$, and similarly each of the $K_{\textrm{ab}}-1$ cavity probabilities $q_{l\rightarrow i}$ should be sampled from the population $Q_{\textrm{ab}}^1$ only with probability $q$ and it should be sampled from the array $Q_{\textrm{ab}}^0$ with the remaining probability $1-q$. After all these $K-1$ input cavity probabilities are sampled, then we again apply Eq.~(\ref{eq:newqij}) to generate a new cavity probability (say $q_{i\rightarrow j}^\prime$) and replace a randomly chosen element of the population array $Q_{\textrm{ba}}^1$ with this newly generated value. We mark the difference between the two newly generated cavity probabilities $q_{i\rightarrow j}$ and $q_{i\rightarrow j}^\prime$ by the following difference measure:
\begin{equation}
\Delta \, = \, \bigl| \ln q_{i\rightarrow j}  - \ln q_{i\rightarrow j}^\prime \bigr| \; .
\end{equation}

The population arrays $Q_{\textrm{bb}}^0$ and $Q_{\textrm{bb}}^1$ can be updated following the same recipe (the only difference is that vertex $i$ has $K_{\textrm{ba}}$ adjacent vertices $k$ in group $A$ and $K_{\textrm{bb}} - 1$ adjacent vertices $l$ in group $B$). The population arrays $Q_{\textrm{ab}}^0$ and $Q_{\textrm{ab}}^1$ are also updated together and by the same recipe, only that all the $K-1$ adjacent vertices $k$ are in group $B$.
  
The averaged value of $\Delta$ is recorded for each unit time of population dynamics. Figure~\ref{fig:K541B2p9:delta} shows some trajectories of $\Delta$ obtained by independent runs of the population dynamics on the graph ensemble with degree parameters $K=5$ and $K_{\textrm{ba}}=4$, at fixed inverse temperature $\beta = 2.9$. We see that $\Delta$ will eventually decay to zero, but the exact time $t_{decay}$ of decay is a random variable. At $\beta = 2.9$, when $\Delta$ decays to zero, we find that the final values of cavity messages $q_{i\rightarrow j}$ of all the six population arrays are all identical to the BP fixed-point value of the ferromagnetic CP phase, suggesting that the CP phase is the globally stable phase at $\beta = 2.9$.  The distributions of this decay time $t_{decay}$ for several different population sizes $\mathcal{P}$ are shown in Fig.~\ref{fig:K541B2p9:tau}. We see that the probability profile of $t_{decay}$ is shifted to the right as $\mathcal{P}$ increases and it becomes more and more skewed. We show the median decay time as a function of population size $\mathcal{P}$ in Fig.~\ref{fig:K541B2p9:scaling}. The median decay time increases superlinearly with $\log \mathcal{P}$. We expect that at the limit of $\mathcal{P}\rightarrow \infty$ the median decay time will diverge to infinity.

Our population dynamics results confirm the existence of many spin glass macroscopic states at $\beta = 2.9$ within the configuration subspace of the paramagnetic DS phase of the planted ($K=5$, $K_{\textrm{ba}}=4$) regular random graph ensemble. This is consistent with the prediction that, for the fully regular random regular graph ensemble with degree $K = 5$, the spin glass dynamical phase transition point is located at inverse temperature $\beta = 2.7726$~\cite{Zhang-Zeng-Zhou-2009}. 

When the vertex degree $K > 16$, the spin glass dynamical phase transition point $\beta_d$ no longer coincides with the spin glass local stability point~\cite{Zhang-Zeng-Zhou-2009}. This fact can also be checked for the planted regular random graph ensembles by performing the same population dynamics simulations.

\end{appendix}

%


%

\end{document}